\documentclass{aa} 
\usepackage{txfonts}
\usepackage{longtable}  
\usepackage{rotating}
\usepackage{natbib}
\usepackage{graphicx}
\usepackage{graphics}
\usepackage{psfrag}
\usepackage{amssymb}
\bibpunct{(}{)}{;}{a}{}{,}
\def\Teff{$T_{\mathrm{eff}}$}
\def\logg{\ensuremath{\log g}}
\def\vmic{$\upsilon_{\mathrm{mic}}$}

\def\vsini{\ensuremath{{\upsilon}\sin i}}
\def\kms{$\mathrm{km\,s}^{-1}$}

\def\vr{${\upsilon}_{\mathrm{r}}$}
\def\logt{\ensuremath{\log t}}
\def\espa{ESPaDOnS}

\def\llm{{\sc LLmodels}}

\def\logl{\ensuremath{\log L/L_{\odot}}}
\def\mbol{$M_{\mathrm{bol}}$}
\begin{document} 
\title{The effect of rotation on the abundances of the chemical elements
of the A--type stars in the Praesepe cluster\footnote{Based on observations made
at the Observatoire de Haute--Provence}} 
\subtitle{} 
\author{L. Fossati\inst{1}       \and 
	S. Bagnulo\inst{2}       \and 
	J. Landstreet\inst{4}    \and 
	G. Wade\inst{5}          \and 
	O. Kochukhov\inst{6}     \and 
	R. Monier\inst{7}        \and 
	W. Weiss\inst{1}         \and 
	M. Gebran\inst{8}} 
\offprints{L.~Fossati} 
\institute{
	Institut f\"ur Astronomie, Universit\"{a}t Wien, 
	T\"{u}rkenschanzstrasse 17, 1180 Wien, Austria.\\
	\email{fossati@astro.univie.ac.at; weiss@astro.univie.ac.at} 
	\and 
	Armagh Observatory, College Hill, Armagh BT61 9DG, Northern Ireland.
	\email{sba@arm.ac.uk} 
	\and 
	Department of Physics \& Astronomy, University of Western Ontario, 
	London, N6A 3K7, Ontario, Canada.\\
	\email{jlandstr@astro.uwo.ca}
	\and  
	Physics Dept., Royal Military College of Canada, PO Box 17000, 
	Station Forces, K7K 4B4, Kingston, Canada.
	\email{Gregg.Wade@rmc.ca}
	\and 
	Department of Physics and Astronomy, Uppsala University, 
	751 20, Uppsala, Sweden. 
	\email{oleg@astro.uu.se}
	\and 
	Laboratoire Universitaire d'Astrophysique de Nice, Universite de 
	Nice Sophia--Antipolis, Parc Valrose, 06000 Nice, France.
	\email{Richard.MONIER@unice.fr}
	\and
	Groupe de Recherche en Astronomie et Astrophysique du Languedoc, 
	UMR  5024, Universit\'e Montpellier II, Place Eug\`ene Bataillon, 
	34095  Montpellier, France. 
	\email{Marwan.Gebran@graal.univ-montp2.fr}
} 
\date{} 
\abstract 
{}
{We study how chemical abundances of late B--, A-- and early F--type stars
evolve with time, and we search for correlations between the abundance of 
chemical elements and other stellar parameters, such as effective temperature 
and \vsini.} 
{We have observed a large number of B--, A-- and F--type stars belonging to 
open clusters of different ages. In this paper we concentrate on the Praesepe 
cluster (\logt\, = 8.85), for which we have obtained high resolution, high 
signal--to--noise ratio spectra of sixteen normal A-- and F--type stars and 
one Am star, using the SOPHIE spectrograph of the Observatoire de 
Haute--Provence. For all the observed stars, we have derived fundamental 
parameters and chemical abundances. In addition, we discuss another 
eight Am stars belonging to the same cluster, for which the abundance analysis 
had been presented in a previous paper.} 
{We find a strong correlation between peculiarity of Am stars and \vsini. 
The abundance of the elements underabundant in Am stars increases with \vsini, 
while it decreases for the overabundant elements. Chemical abundances of 
various elements appear correlated with the iron abundance.} 
{} 
\keywords{}
\titlerunning{Effect of rotation in abundances of A--type stars in the 
Praesepe cluster}
\authorrunning{L.~Fossati et al.}
\maketitle
\section{Introduction}\label{introduction}
In stellar astrophysics, the study of the atmospheres of A-- and B--type stars 
plays a very special role. The atmospheres of these stars display 
a variety of different phenomena, such as the presence of large and 
relatively simple magnetic fields, strong surface convection, pulsation, 
diffusion of chemical elements, and various kinds of mixing processes from 
small--scale turbulence to global circulation currents. These general 
physical processes are almost certainly active in the atmospheres (and 
interiors) of stars other than the main sequence A-- and B--type 
stars, but in more subtle ways. Because of the easily visible effects found 
in A-- and B--type stars, these stars provide unique access to the invisible
interior processes. To fully understand the actual role of all these physical 
phenomena, it is very important to seek constraints from the observations, 
in particular to perform a detailed study of a large number of stars of
different ages and peculiarity types.

For this project, it is very interesting to study stars that are cluster 
members, rather than selecting targets in the field, because of two very 
compelling reasons. First, we can assume that all cluster members have 
approximately the same original chemical composition and age. Therefore, 
when we analyse the chemical abundances of stars belonging to the same 
cluster, we can assume that we are studying objects that are different only 
by their initial mass, rotational velocity, magnetic field strength and 
binarity. Second, the age of stars 
belonging to open clusters can be determined with much higher accuracy than 
for objects in the field, especially for stars that are young enough that
less than half of their life in the main sequence has elapsed
\citep[for a discussion of the star's age determination see][]{stefano2006}. 
Therefore, we have carried out two large observational campaigns. With FORS1 
at the ESO VLT and with \espa\, at the Canada--France--Hawaii Telescope (CHFT) 
we have made a survey of magnetic A-- and B--type stars in open clusters, to 
search for links between magnetic fields and stellar evolution. The results 
of this campaign are described by \citet{stefano2006,john2006,john2007}. 
In the framework of this observational campaign, we have also obtained 
high--resolution spectroscopy of a large number of early--type stars in about 
ten open clusters of different ages. The clusters are approximately 
uniformly distributed in logarithmic age from \logt\, = 6.8 to \logt\, = 8.9, 
and the spectra have been obtained using five different spectrographs: 
FLAMES at the ESO VLT, FIES at the Nordic Optical Telescope (NOT), ELODIE 
and SOPHIE at the Observatoire de Haute--Provence (OHP), and \espa\, at the 
CFHT.

In a previous paper \citep[][hereafter referred to as Paper~I]{luca}, we 
started the analysis of the Praesepe cluster, presenting the observations and 
the results of the abundance analysis of eight Am and three normal A-- and 
F--type stars. In this work we complete the study of the Praesepe cluster 
by presenting observations of 16 additional normal A-- and early F--type 
stars, and an additional Am star. These new observations were obtained with 
the SOPHIE spectrograph at the Observatoire de Haute--Provence. Because of the
large sample of analysed stars, we are able to make a sensitive search for 
correlations between the abundances of the chemical elements and other stellar 
parameters, such as  the effective temperature \Teff\, and projected rotational 
velocity \vsini.

This Paper is organised as follows. In Sect.~\ref{target selection} we 
describe the criteria that we have adopted for target selection, which were 
used both for the entire observational campaign, and for the specific 
observations of the Praesepe cluster presented in this paper. Details about 
observations of the Praesepe cluster and data reduction are given in 
Sect.~\ref{observations}. In Sect.~\ref{SOPHIEESP} we compare spectra of 
a reference star obtained with two different instruments that we 
have used in our observational campaign: the SOPHIE spectrograph and the 
\espa\, spectropolarimeter. In Sect.~\ref{spectral analysis} we describe the
procedure used to perform the abundance analysis, and we present the results 
obtained for the early--type stars of the Praesepe cluster observed 
with the SOPHIE spectrograph. In Sect.~\ref{discussion} we consider the 
abundances of chemical elements for the stars studied in Paper~I and for 
the stars studied in this work. We search for correlations between chemical 
element abundances and effective temperature, \vsini, M/M$_\odot$ and 
fractional age ($\tau$ -- fraction of main sequence lifetime completed). In 
Sect.~\ref{conclusions} we summarise our conclusions. 
\section{Target selection}\label{target selection}
For target selection in the overall program, we assigned the highest priority 
to the most probable cluster members that were already known to be chemically 
peculiar (Am, Ap, HgMn stars) and/or to show slow rotation (typically 
\vsini\, $<$ 50 \kms). For this purpose we have followed a standard 
procedure to select the stars to observe and to assign their priority. 

The first step was to identify the most probable cluster members by checking 
the membership probabilities given by \citet{robichon} and \citet{baumgardt} 
obtained from HIPPARCOS data, by \citet{kharchenko2004} and by 
\citet{dias2006}. We considered as probable members of the cluster every star 
with a kinematic and photometric membership probability higher than 0.14 
(Kharchenko, private communication) in the case of the \citet{kharchenko2004} 
catalogue, or a membership probability higher than 10\% for the 
\citet{dias2006} catalogue. 

We then searched for information about the spectral type, magnitude, \vsini, 
binarity, variability, chemical peculiarities, magnetic field and radial
velocity for each star considered a probable member of the cluster. We 
collected information from 
SIMBAD\footnote{\tt http://simbad.u-strasbg.fr/simbad/sim-fid} (scanning also 
the different references mentioned therein), 
VIZIER\footnote{\tt http://vizier.u-strasbg.fr/viz-bin/VizieR} (using a 
combination of several keywords -- binaries:cataclysmic, binaries:eclipsing, 
binaries:spectroscopic, multiple\_stars, open\_clusters, positional\_data, 
proper\_motions, rotational\_velocities, spectral\_classification, 
spectrophotometry, spectroscopy, stars:early--type, stars:late--type, 
stars:peculiar, stars:variable -- in a range of 10 arcsec around the position 
of each star) and the WEBDA\footnote{\tt http://www.univie.ac.at/webda/}
database \citep{webda}. We have also looked for additional information about 
each star in articles by \citet{glebocki} and \citet{royer} for \vsini\, 
measurements, \citet{rodriguez} for $\delta$Sct pulsations, the SB9 catalogue 
\citep{SB9catalogue} for information about a possible binarity and 
\citet{renson} for chemical peculiarities. 

We have then searched for archival high resolution spectra in several archives
(e.g. the ESO archive\footnote{\tt http://archive.eso.org/} and the ELODIE
archive\footnote{\tt http://atlas.obs-hp.fr/elodie/}), in order to avoid a 
duplication of observational effort, in case some stars had already been 
observed with high resolution spectroscopy. Almost no pre--existing spectra 
were found.

Of the sample obtained with this procedure, we kept only the stars with 
spectral types between F5 (to avoid too long an exposure time on a single 
object) and B5 (to avoid strong stellar winds), giving a higher rank to 
peculiar stars and slow rotators. From the sample we have removed stars 
already known to be SB2 systems, unless they had already been extensively 
observed using high--resolution spectroscopy.
\section{Observations and data reduction}\label{observations}
We have observed seventeen stars of the Praesepe cluster using the SOPHIE 
spectrograph at the Observatoire de Haute--Provence (OHP) during two runs,
from December 1st to 2nd 2006 and from 
March 10th to 12th 2007. 

SOPHIE is a cross-dispersed \'{e}chelle spectrograph mounted at the 1.93--m 
telescope at the OHP. The spectrograph is fed from the Cassegrain focus 
through a pair of optical fibers, one of which is used for starlight and the 
other can be used for either the sky background or the wavelength calibration 
lamp, but can also be masked. The spectra cover the wavelength range 
3872--6943~\AA. The instrument allows observations either with medium spectral 
resolution (R$\sim$40\,000) or with high spectral resolution (R$\sim$75\,000). 
Since for most of the stars of our sample no precise \vsini\, measurements were 
available, we decided to observe all of them in high resolution mode.

The spectra were automatically reduced using the SOPHIE pipeline, adapted from 
the HARPS software designed by Geneva Observatory. A detailed description of 
the pipeline is available on the SOPHIE web page 
\footnote{\tt http://www.obs-hp.fr/www/guide/sophie/sophie-eng.html}.

From our spectra we found that almost all of the stars have high rotational 
velocity (\vsini\, $\geq$ 30 \kms). For these objects the continuum 
normalisation is a critical reduction procedure that cannot be performed with 
an automatic fitting procedure. We therefore rectified the continua of these 
stars manually. Even with manual rectification, it was not possible to 
determine a correct continuum level shortwards of the H$\gamma$ line 
(4340.462~\AA), since there were not enough continuum windows in the spectra, 
due to the crowding of spectral lines when \vsini\, is high. 

It is well known how important it is to reach a high signal--to--noise ratio 
(SNR) to be able to carry out an abundance analysis of rapidly rotating stars 
\citep{hill}. For two stars (HD~73798 and HD~73993) the SNR was too low, so we 
have rebinned the spectra to increase the SNR, at the expense of 
spectral resolution.

The complete sample of stars observed and analysed in this paper is listed in
Table~\ref{tabella radec}. In the sample, one object is an Am star, while the
others are normal stars with spectral types between F2 and A1. Seven stars 
of our sample are present in the $\delta$Sct variable star catalog of 
\citet{rodriguez}. One star (HD~73993) is present in the Combined General 
Catalogue of Variable Stars \citep{samus}, but the variability type is 
listed as "unknown". HD~73666 is the only known binary star present in our 
sample, as we have removed all SB2 stars from the selected target 
stars, as explained in Sect.~\ref{target selection}. 
\begin{table*}[ht]
\caption[ ]{Basic data of the observations for the program stars. The SNR are 
calculated at $\sim$5500~\AA\ in a bin of 0.5 \AA. 
The exposure time is in seconds. The HJD indicate the Heliocentric Julian Date at the middle of 
the exposure. The $\delta$Sct stars are present in the catalog of variable stars by 
\citet{rodriguez}. $^{*}$: the star is included, as variable of an "unknown" type, in the 
Combined General Catalogue of Variable Stars by \citet{samus}.}
\label{tabella radec}
\centering                      
\begin{tabular}{ccccccc}
\hline
\hline
HD & HJD & M$_{\it{v}}$ & Spectral Type & SNR & Exp. Time & Remarks \\
\hline
72757 & 2454172.336 & 8.45 & F0   & 190 & 3600   &		    \\
72846 & 2454070.737 & 7.48 & A5V  & 210 & 3600   &		    \\
73175 & 2454171.406 & 8.25 & F0Vn & 155 & 3600   & $\delta$Sct      \\
73345 & 2454171.363 & 8.14 & A7V  & 188 & 3600   & $\delta$Sct      \\
73450 & 2454171.494 & 8.59 & A9V  & 185 & 3600   & $\delta$Sct      \\
73574 & 2454171.585 & 7.75 & A8V  & 170 & 3600   &		    \\
73666 & 2454070.682 & 6.61 & A1V  & 326 & 1800   & speckle binary   \\
73746 & 2454172.514 & 8.72 & F0V  & 154 & 3600   & $\delta$Sct      \\
73798 & 2454172.424 & 8.48 & F0V  & 190 & 3600   & $\delta$Sct      \\
73993 & 2454171.450 & 8.56 & F2V  & 134 & 3600   & variable$^{*}$   \\
74028 & 2454171.318 & 7.97 & A7V  & 185 & 3600   & $\delta$Sct      \\
74050 & 2454170.566 & 7.91 & A6Vn & 160 & 3600   & $\delta$Sct      \\
74135 & 2454172.559 & 8.88 & F0   & 120 & 3600   &		    \\
74587 & 2454172.468 & 8.51 & A5   & 177 & 3600   &		    \\
74589 & 2454172.379 & 8.46 & F0   & 200 & 3600   &		    \\
74656 & 2454170.537 & 8.04 & Am   & 226 & 2x3600 & sum of two exposures\\
74718 & 2454171.538 & 8.39 & A5   & 172 & 3600   &		    \\
\hline
\end{tabular}
\end{table*}

\section{SOPHIE vs. \espa}\label{SOPHIEESP}
The observational material employed in this large project has been obtained 
with different instruments at different telescopes. Since the homogeneity of 
the analysis is of crucial importance, in this section we compare spectra of
a reference star obtained with two different spectrographs to demonstrate 
the homogeneity of our data.

HD~73666 was previously analysed in Paper~I. The star was also used as
reference star by \citet{tanya-reference}. Here we compare the spectra of
HD~73666 obtained with \espa\, (analysed in Paper~I) and with SOPHIE. \espa\, 
was mounted at the Canada-France-Hawaii Telescope (CFHT) in 2005 and SOPHIE 
at the Observatoire de Haute--Provence (OHP) in 2006. The two instruments have 
similar mean resolving power, 65\,000 and 75\,000 respectively. \espa\, is 
also adapted for spectropolarimetric observations, while SOPHIE for radial 
velocity observations. Here we compare the spectra of the reference star 
obtained with the two instruments.

\begin{figure}[ht]
\begin{center}
\includegraphics[width=95mm]{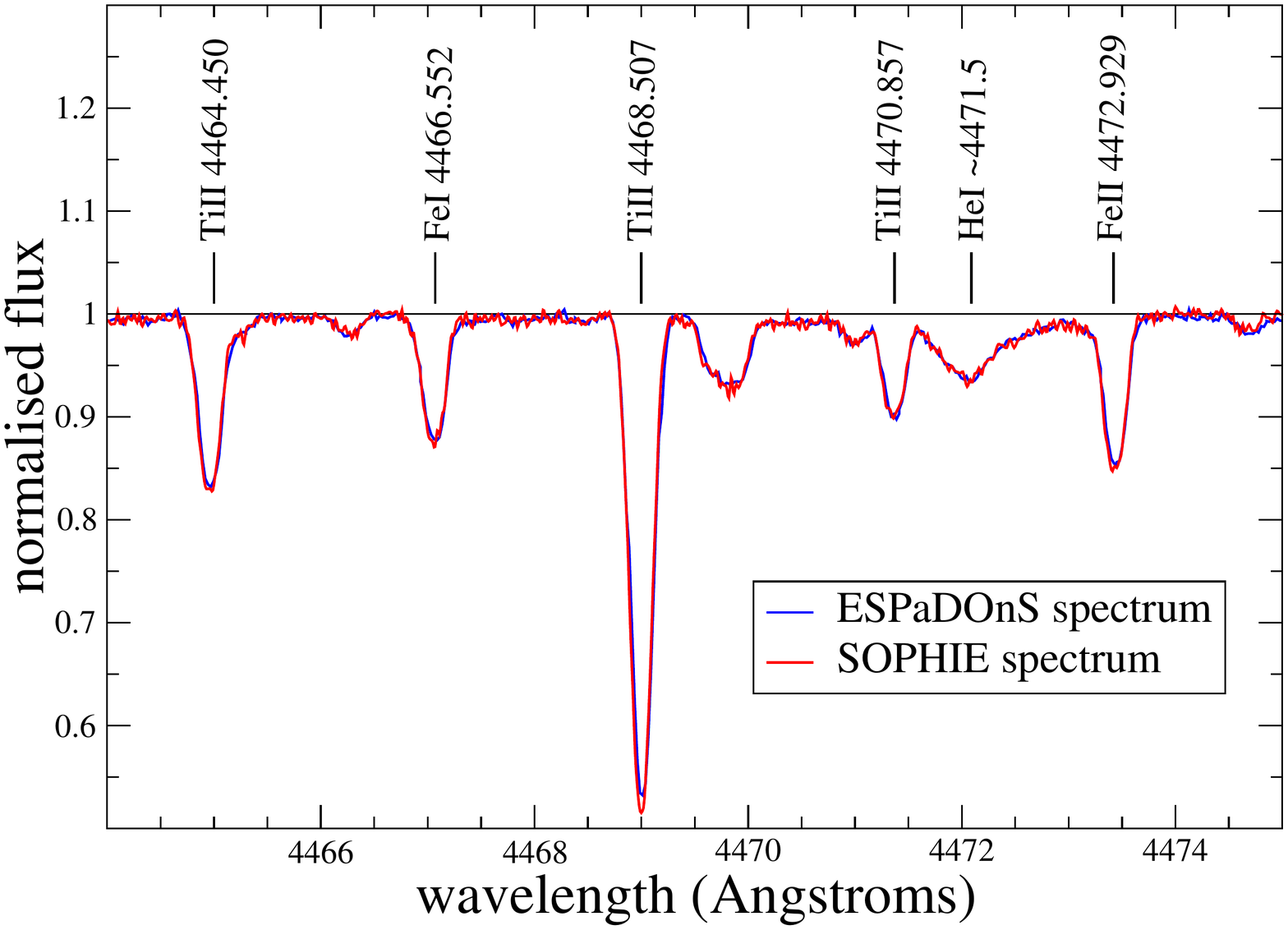}
\caption{Comparison between the spectra of HD~73666 obtained with the \espa\,
spectropolarimeter (blue line) and the SOPHIE spectrograph (red line) in the
region of the \ion{He}{i} multiplet at $\sim$4471.5~\AA. No difference is 
visible between the two spectra, apart the one due to the different spectral 
resolution of the two instruments.} 
\label{comparisonHe} 
\end{center} 
\end{figure}

\begin{figure}[ht]
\begin{center}
\includegraphics[width=95mm]{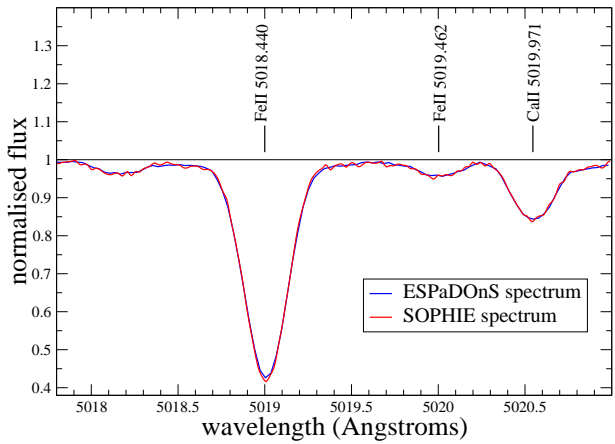}
\caption{Same as for Fig.~\ref{comparisonHe}, but in the region of the strong
\ion{Fe}{ii} line at 5018.440~\AA.} 
\label{comparison5018} 
\end{center} 
\end{figure}

In Fig.~\ref{comparisonHe} and Fig.~\ref{comparison5018} (available online) we 
show a direct comparison between the \espa\, and SOPHIE spectra for HD~73666 
around the He multiplet at $\sim$4471.5~\AA\, and the strong \ion{Fe}{ii} line 
at 5018.440~\AA\, respectively. No significant differences are present apart 
from those due to the small difference of spectral resolution, visible mainly 
in the cores of the strong lines.

HD~73666 was first detected as a binary star with speckle interferometry by
\citet{hartkopf1984}, and was observed and analysed further by 
\citet{mcalister}. \citet{mason1993}, who collected all available 
interferometric measurements and concluded that HD~73666 "shows little orbital 
motion". Radial velocity measurements are reported by \citet{abt1970}, 
\citet{conti1974}, \citet{abt1999} and \citet{soren2002}. All these authors 
give radial velocities compatible with the cluster mean of 
34.5$\pm$0.0 \kms\, \citep{robichon}. The spectra plotted in 
Fig.~\ref{comparisonHe} and Fig.~\ref{comparison5018} have been corrected only 
for the terrestrial radial velocity; nevertheless, no radial velocity 
difference is visible from the comparison of the two spectra, in agreement 
with results from previous authors.
\section{Model atmosphere and abundance analysis}
\label{spectral analysis}
Model atmospheres were calculated with the LTE code \llm, which uses direct 
sampling of the line opacities \citep{denis2004} and makes it possible to 
compute model atmospheres with an individualised abundance pattern. We adopted 
the VALD database \citep{vald1,vald2,vald3} as the source of spectral line data, 
including lines originating from predicted levels. We have carried out a 
preselection procedure to eliminate all lines that do not contribute 
significantly to the line opacity. This was done by requiring that the 
line--to--continuum opacity ratio at the center of each included line be 
greater then 0.05~\%. Convection was treated according to the CM approach 
\citep{CM}. 

The initial values of the fundamental atmospheric parameters (\Teff\, and 
\logg) were derived from Str\"omgren \citep{hauck} and Geneva \citep{webda} 
photometry, using calibrations from \citet{napiwotzki} and
\citet{geneva_cal} respectively.

The adopted \Teff\, and \logg\, were derived spectroscopically, as explained 
in detail in Paper~I. We have almost always used Fe lines for this purpose, 
as the generally high \vsini\, of the analysed stars did not provide a suitable 
number and variety of lines of other elements. The microturbulent velocity 
(\vmic) was determined spectroscopically, from Fe and \ion{Ti}{ii} lines, 
for HD~74656 only; this was possible because of its low \vsini. For all other 
stars we have adopted \vmic\, computed from the relation \citep{pace2006}:
\begin{equation}
\label{vmicphotometry}
\upsilon_{\mathrm{mic}}=-4.7\log(T_{\mathrm{eff}})+20.9\,\,\mathrm{km\,s}^{-1}\ .
\end{equation}
There is no significant difference between adopting the \vmic\, given by 
Eq.~\ref{vmicphotometry} and assuming a fixed value of 2.7 \kms. We notice 
that \vmic\, obtained for the Am star HD~74656 is significantly larger than
2.7 \kms. Whether this is also the case for other stars of the sample cannot
be tested with the available data.

Radial velocity (\vr) and \vsini, in \kms, were determined by computing the 
median of the results obtained by synthetic fitting of several individual 
lines in the observed spectrum. We have derived error bars for these 
quantities for each analysed star using the standard deviation of the derived 
values. The error bar associated with \vr\, is strongly dependent on \vsini, 
ranging from about 1 \kms\, for slow rotators, up to about 8 \kms\, for the 
fast rotators. The error bars in \vsini\, are about 5~\% or 3~\kms, whichever 
is larger. 

Table~\ref{parameters} reports the fundamental parameters obtained for 
each star of the sample from photometry and spectroscopy. The parameters
adopted for abundance analysis are those obtained from spectroscopy.
\begin{table*}[ht]
\caption[ ]{Atmospheric parameters for the analysed stars of the Praesepe cluster. 
Columns two to five show the fundamental parameters derived from Str\"{o}mgren and Geneva photometry.
The sixth and seventh columns show the final adopted parameters derived from spectroscopy.
The errors on the fundamental parameters are estimated to be 200\,K, 0.7\,\kms and 0.2 dex for \Teff,
\vmic\, and \logg\, respectively. The estimated error on \vsini\, is about 5\%.
\vr\, is given in \kms. The uncertainty in \vr\, depends strongly on the \vsini. For a low \vsini\, value 
the error bar on \vr\, is estimated to be about 1 \kms, that increases till about 8 \kms\, for the faster rotators.}
\label{parameters}
\centering                      
\begin{tabular}{cccccccccccc}
\hline
\hline
 & \multicolumn{2}{c}{Str\"{o}mgren photometry} & \multicolumn{2}{c}{Geneva photometry} & \multicolumn{2}{c}{final set} & & & & & \\
\hline
HD & \Teff & \logg  & \Teff & \logg & \Teff & \logg & \vmic  & \vsini & $\sigma_{\ensuremath{{\upsilon}\sin i}}$ & \vr & $\sigma_{{\upsilon}_{\mathrm{r}}}$ \\
   &   [K] & [cgs]  &  [K]  & [cgs] &   [K] & [cgs] & [\kms] & [\kms] & [\kms] & [\kms] & [\kms] \\
\hline											
72757 & 7330 & 3.81 &      &      & 7400 & 3.60 & 2.7 & 179 & 10 & 35.8 & 7.0 \\	
72846 & 8187 & 3.85 & 8146 & 4.12 & 8045 & 3.50 & 2.5 & 119 &  6 & 35.1 & 7.3 \\	
73175 & 7735 & 3.94 & 7506 & 4.22 & 7660 & 3.94 & 2.6 & 163 &  9 & 36.7 & 5.7 \\	
73345 & 7779 & 3.92 & 7865 & 4.43 & 7993 & 3.96 & 2.6 & 85  &  4 & 34.5 & 4.4 \\	
73450 & 7271 & 3.77 & 7455 & 4.26 & 7270 & 4.20 & 2.7 & 138 &  7 & 34.5 & 5.1 \\	
73574 & 7659 & 3.90 & 7733 & 4.36 & 7662 & 4.00 & 2.6 & 102 &  4 & 35.3 & 4.5 \\	
73746 & 7390 & 4.01 & 7236 & 4.16 & 7440 & 4.08 & 2.7 & 95  &  4 & 34.2 & 3.6 \\	
73798 & 7355 & 3.80 & 7301 & 4.11 & 7328 & 3.95 & 2.7 & 200 & 11 & 34.1 & 5.9 \\	
73993 & 7124 & 3.74 & 7153 & 4.11 & 7138 & 3.92 & 2.8 & 240 & 14 & 33.6 & 7.5 \\	
74028 & 7517 & 3.65 & 7772 & 4.34 & 7750 & 4.50 & 2.6 & 150 &  9 & 35.3 & 7.4 \\	
74050 & 7767 & 3.78 & 7722 & 4.19 & 7872 & 3.66 & 2.6 & 188 &  8 & 36.0 & 7.8 \\	
74135 &      &      & 7016 & 4.22 & 7400 & 4.00 & 2.7 & 100 &  4 & 32.5 & 5.7 \\	
74587 &      &      & 7246 & 4.15 & 7500 & 4.20 & 2.7 & 90  &  4 & 33.9 & 4.1 \\	
74589 & 7581 & 4.04 &      &      & 7550 & 4.03 & 2.7 & 127 &  5 & 34.1 & 4.7 \\	
74656 &      &      & 7517 & 4.11 & 7800 & 3.99 & 3.6 & 25  &  1 & 34.1 & 1.3 \\	
74718 & 7599 & 3.99 & 7453 & 4.21 & 7600 & 4.00 & 2.7 & 155 &  7 & 34.1 & 5.7 \\	
\hline
\end{tabular}
\end{table*}



\begin{figure}[ht]
\begin{center}
\includegraphics[width=90mm]{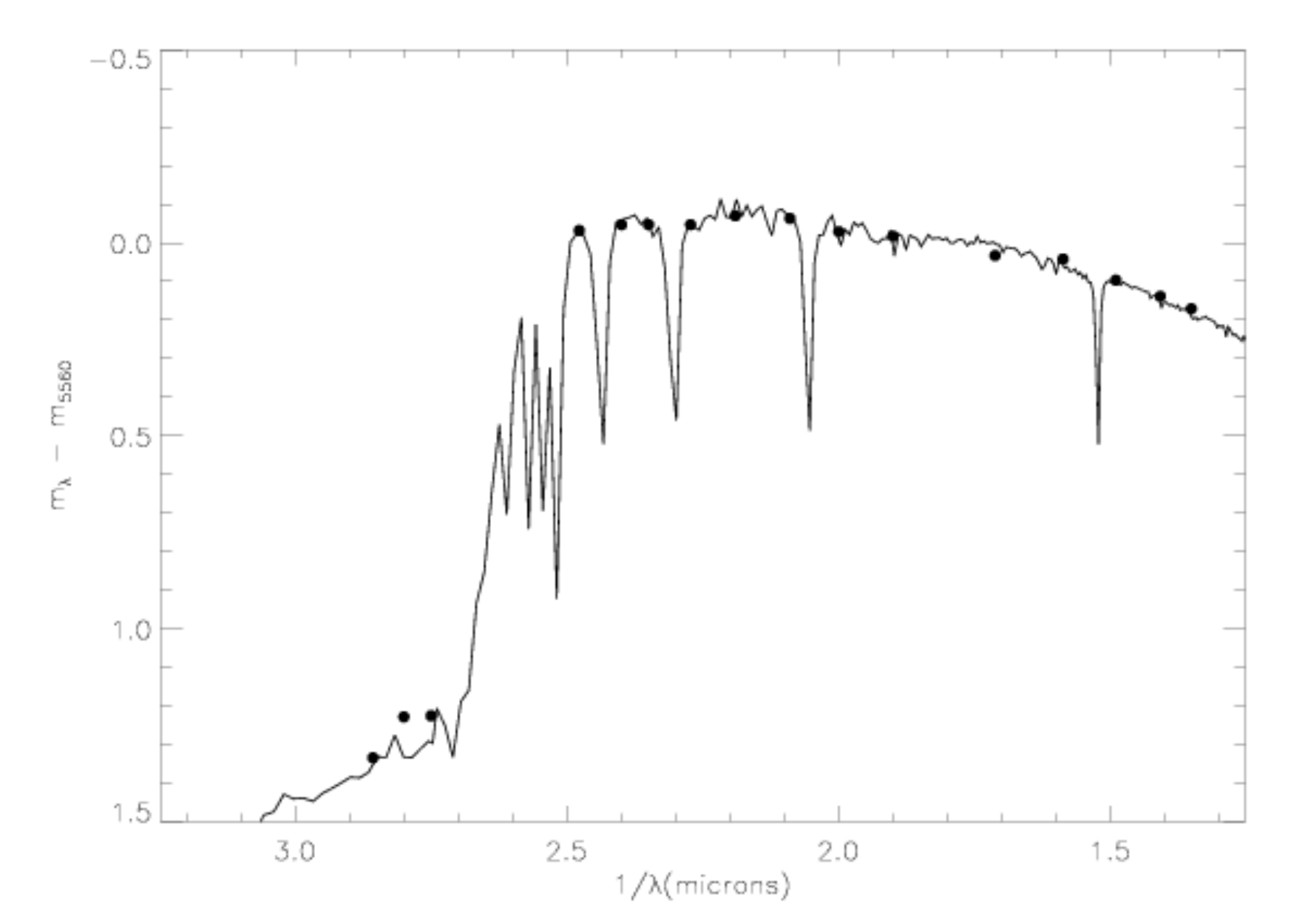}
\caption{Comparison between spectrophotometry from \citet{clampitt} and
normalised fluxes at 5560~\AA\, with the fundamental parameters adopted for the
abundance analysis of the star HD~73345.} 
\label{spectrophotometry} 
\end{center} 
\end{figure}

We have tested the fundamental parameters adopted for the abundance analysis
with spectrophotometry from \citet{clampitt} for 5 stars of the sample. We 
used Atlas9 models and the fundamental parameters adopted for the 
abundance analysis to compute the fluxes for comparison with the 
spectrophotometry. The fluxes were normalised to the flux at 5560~\AA. An 
example of the comparison is shown in Fig.~\ref{spectrophotometry}. We have 
found good agreement, within the error bars of the fundamental parameters, for 
all the stars for which spectrophotometry was available. 

The synthetic line spectrum was produced with the synthesis code Synth3 
\citep{synth3}. In each spectrum we have fit the cores of selected lines of 
each ion to obtain a value of the abundance associated with each line. The 
adopted abundance of that ion is then defined to be the mean of the 
abundances obtained from the selected lines. The error bars associated with 
the mean abundances are the standard deviations, and they do not include 
uncertainties in fundamental parameters (see below). The abundances and their 
standard error bars are given in Table~\ref{abundance_table}. 
\begin{table*}[ht]
\caption[ ]{Abundances ($\log(N_{X}/N_{tot})$) of the program stars with the 
estimated internal errors in units of 0.01 dex, in parenthesis. The internal 
error associated with the derived abundance of each element is the 
standard deviation from the mean abundance of the selected lines of that 
element. For comparison, the solar abundances \citep{asplundetal2005} are given in the 
last column. At.N. gives the Atomic Numbers of the elements. Abundances obtained 
from just one line have no error ($-$;1). Upper limits are denoted by $<$. The second 
number in the brackets is the number of lines used to derive the mean abundance.}
\label{abundance_table}
\scriptsize{
\begin{tabular}{cccccccccccccc}
\hline
\hline
\multicolumn{2}{c}{ } & \multicolumn{8}{c}{"Normal" A$-$type stars} & Solar \\
At.N.&Element& HD~72846    & HD~73345     & HD~73450     & HD~73574     & HD~74028     & HD~74050     & HD~74587     & HD~74718 & Abundances \\
\hline
3 & Li      &$<-8.08(-;1) $&$<-8.33(-;1) $&$<-8.70(-;1) $&$<-8.38(-;1) $&              &              &$<-8.41(-;1) $&$<-8.26(-;1) $&$-10.99$\\  
6 & C       &$-3.58(-;1)  $&$-3.44(12;3) $&$-3.27(-;1)  $&$-3.36(18;2) $&$-3.39(08;2) $&$-3.52(-;1)  $&$-3.49(01;2) $&$-3.51(04;2) $&$-3.65 $\\  
8 & O       &$-3.18(-;1)  $&$-3.22(01;2) $&              &              &              &$-3.70(-;1)  $&$-3.30(-;1)  $&              &$-3.38 $\\  
11& Na      &$-5.44(01;2) $&$-5.37(01;2) $&$-6.28(-;1)  $&$-5.57(02;2) $&$-5.98(-;1)  $&$-5.64(13;2) $&$-5.61(02;2) $&$-5.70(14;2) $&$-5.87 $\\  
12& Mg      &$-4.18(08;3) $&$-4.18(02;3) $&$-5.02(18;2) $&$-4.37(04;3) $&$-4.86(08;3) $&$-4.22(05;4) $&$-4.56(08;3) $&$-4.52(01;2) $&$-4.51 $\\  
14& Si      &$-4.62(16;2) $&$-4.67(-;1)  $&$-4.13(-;1)  $&$-4.19(-;1)  $&$-4.17(-;1)  $&$-4.37(-;1)  $&$-4.16(-;1)  $&$-4.25(-;1)  $&$-4.53 $\\  
16& S       &$-4.71(04;2) $&$-4.44(03;4) $&$-4.35(-;1)  $&$-4.61(02;2) $&$-4.26(01;2) $&              &$-4.50(04;2) $&$-4.28(11;2) $&$-4.90 $\\  
20& Ca      &$-5.17(-;1)  $&$-5.39(09;6) $&$-5.95(06;4) $&$-5.86(16;5) $&$-5.37(16;2) $&$-6.13(06;2) $&$-5.49(15;6) $&$-5.68(02;3) $&$-5.73 $\\  
21& Sc      &$-8.88(-;1)  $&$-8.63(07;3) $&$-8.57(14;3) $&$-8.89(02;3) $&$-8.35(-;1)  $&$-8.96(27;3) $&$-8.56(-;1)  $&$-8.69(14;2) $&$-8.99 $\\  
22& Ti      &$-6.88(03;5) $&$-6.95(06;6) $&$-7.30(11;5) $&$-6.98(09;5) $&$-6.78(-;1)  $&$-7.08(15;5) $&$-6.83(16;3) $&$-6.93(10;5) $&$-7.14 $\\  
24& Cr      &$-6.23(06;3) $&$-6.22(08;2) $&$-6.56(08;3) $&$-6.19(16;3) $&$-6.23(12;4) $&$-6.48(10;3) $&$-6.05(13;4) $&$-6.44(20;5) $&$-6.40 $\\  
25& Mn      &              &$-6.37(-;1)  $&$-6.88(-;1)  $&$-6.52(02;2) $&$-6.77(-;1)  $&$-6.61(-;1)  $&$-6.62(04;2) $&$-6.71(-;1)  $&$-6.65 $\\  
26& Fe      &$-4.55(18;42)$&$-4.33(11;61)$&$-4.62(09;15)$&$-4.49(10;30)$&$-4.50(09;18)$&$-4.44(13;16)$&$-4.28(10;33)$&$-4.61(11;26)$&$-4.59 $\\  
28& Ni      &$-5.70(18;2) $&$-5.58(11;4) $&$-5.82(16;2) $&$-5.62(08;4) $&$-5.93(14;3) $&$-5.60(15;3) $&$-5.84(-;1)  $&$-5.68(02;3) $&$-5.81 $\\  
39& Y       &$-9.75(-;1)  $&$-9.46(-;1)  $&$-9.83(-;1)  $&$-9.20(-;1)  $&$-9.56(-;1)  $&$-9.26(-;1)  $&$-9.13(-;1)  $&$-9.10(-;1)  $&$-9.83 $\\  
56& Ba      &$-9.48(-;1)  $&$-9.30(06;2) $&$-9.50(02;2) $&$-8.98(04;2) $&$-9.65(-;1)  $&$-9.52(01;2) $&$-8.96(25;2) $&$-9.15(-;1)  $&$-9.87 $\\  
\hline
&   \Teff   & 8045	   & 7993         & 7270	 & 7662         & 7750	       & 7872	      & 7500         & 7600	    & \\
&   \logg   & 3.50	   & 3.96         & 4.20	 & 4.00         & 4.50	       & 3.66	      & 4.20         & 4.00	    & \\
&   \vmic   & 2.5	   & 2.6	  & 2.7	         & 2.6          & 2.6	       & 2.6	      & 2.7          & 2.7	    & \\
&   \vsini  & 119	   & 85	          & 138	         & 102          & 150	       & 188	      & 90	     & 155	    & \\
\hline
\hline
\multicolumn{2}{c}{ } & \multicolumn{7}{c}{F$-$type stars} & \multicolumn{1}{c}{Am star} & Solar \\
At.N.&Element& HD~72757    & HD~73175     & HD~73746     & HD~73798	& HD~73993     & HD~74135     & HD~74589     & HD~74656 & Abundances \\
\hline
3 & Li      &              &              &$<-8.70(-;1) $&		&	       &$<-8.47(-;1) $&$<-8.35(-;1) $&$-8.61(-;1)  $&$-10.99$\\  
6 & C	    &$-3.71(16;2) $&$-3.36(01;2) $&$-3.39(-;1)  $&		&	       &$-3.30(14;2) $&$-3.19(18;2) $&$-4.09(-;1)  $&$-3.65 $\\  
7 & N	    &              &              &              &		&	       &  	      &	             &$-3.98(-;1)  $&$-4.26 $\\  
8 & O	    &              &              &$-3.30(-;1)  $&		&	       &$-3.40(-;1)  $&$-3.36(-;1)  $&$-3.88(-;1)  $&$-3.38 $\\  
11& Na      &$-5.61(-;1)  $&$-5.81(01;2) $&$-5.45(13;2) $&		&	       &$-5.73(24;2) $&$-5.73(07;2) $&$-5.35(01;2) $&$-5.87 $\\  
12& Mg      &$-4.56(14;2) $&$-4.38(07;2) $&$-4.50(11;5) $&$-4.76(23;2) $&$-4.57(-;1)  $&$-4.14(09;2) $&$-4.25(02;2) $&$-4.73(11;3) $&$-4.51 $\\  
13& Al      &		   &	          &              &		&	       &  	      &	             &$-5.15(-;1)  $&$-5.67 $\\  
14& Si      &$-4.23(06;3) $&$-4.39(-;1)  $&$-4.29(-;1)  $&$-4.09(-;1)  $&$-4.34(-;1)  $&$-4.14(-;1)  $&$-4.20(-;1)  $&$-4.37(-;1)  $&$-4.53 $\\  
16& S	    &$-4.62(01;2) $&$-4.55(-;1)  $&$-4.54(05;5) $&		&	       &$-4.62(05;2) $&$-4.44(07;2) $&$-4.52(04;5) $&$-4.90 $\\  
20& Ca      &$-5.60(11;6) $&$-5.41(03;3) $&$-5.68(16;8) $&$-5.38(15;5) $&$-5.65(-;1)  $&$-5.45(14;4) $&$-5.53(05;7) $&$-6.34(13;3) $&$-5.73 $\\  
21& Sc      &$-9.07(-;1)  $&$-8.66(-;1)  $&$-8.98(10;3) $&$-9.20(-;1)  $&$-8.90(-;1)  $&$-8.48(15;4) $&$-8.43(25;3) $&$-9.27(-;1)  $&$-8.99 $\\  
22& Ti      &$-7.03(16;5) $&$-6.90(15;3) $&$-6.68(13;3) $&$-6.90(-;1)  $&$-7.18(-;1)  $&$-7.03(09;6) $&$-6.92(11;7) $&$-6.90(16;18)$&$-7.14 $\\  
23& V	    &		   &	          &              &		&	       &  	      &	             &$-7.06(-;1)  $&$-8.04 $\\  
24& Cr      &$-6.72(08;3) $&$-6.55(01;2) $&$-6.20(23;6) $&$-6.64(36;4) $&	       &$-6.16(12;3) $&$-6.26(20;4) $&$-5.72(10;11)$&$-6.40 $\\  
25& Mn      &$-6.92(-;1)  $&$-6.98(08;2) $&$-6.46(01;2) $&$-6.36(-;1)  $&	       &$-6.42(11;3) $&	             &$-6.27(11;5) $&$-6.65 $\\  
26& Fe      &$-4.50(08;31)$&$-4.52(08;14)$&$-4.46(08;49)$&$-4.47(06;17)$&$-4.47(08;6) $&$-4.45(11;60)$&$-4.51(11;53)$&$-4.11(13;74 $&$-4.59 $\\  
27& Co      &		   &	          &  	         &		&	       &  	      &	             &$-5.38(-;1)  $&$-7.12 $\\  
28& Ni      &$-5.57(-;1)  $&$-5.68(05;3) $&$-5.38(12;5) $&$-6.20(07;2) $&$-6.15(-;1)  $&$-5.53(10;4) $&$-5.90(17;4) $&$-5.00(08;15)$&$-5.81 $\\  
29& Cu      &		   &	          &  	         &		&	       &  	      &	             &$-6.92(-;1)  $&$-7.83 $\\  
30& Zn      &		   &	          &  	         &		&	       &  	      &	             &$-6.77(-;1)  $&$-7.44 $\\  
38& Sr      &		   &	          &  	         &		&	       &  	      &	             &$-8.23(-;1)  $&$-9.12 $\\
39& Y	    &$-10.07(-;1) $&$-9.39(-;1)  $&$-9.35(01;2) $&$-9.20(-;1)  $&	       &$-9.46(-;1)  $&$-8.99(-;1)  $&$-8.92(11;3) $&$-9.83 $\\  
40& Zr      &		   &	          &  	         &		&	       &  	      &	             &$-8.62(-;1)  $&$-9.45 $\\  
56& Ba      &$-9.12(06;2) $&$-9.64(-;1)  $&$-9.02(12;2) $&$-8.78(-;1)  $&$-9.48(-;1)  $&$-9.10(-;1)  $&$-9.14(-;1)  $&$-8.46(23;3) $&$-9.87 $\\  
57& La      &		   &	          &  	         &		&	       &  	      &	             &$-9.22(14;3) $&$-10.91$\\  
58& Ce      &		   &	          &  	         &		&	       &  	      &	             &$-9.11(08;3) $&$-10.46$\\  
60& Nd      &		   &	          &  	         &		&	       &  	      &	             &$-9.27(13;5) $&$-10.59$\\  
62& Sm      &		   &	          &  	         &		&	       &  	      &	             &$-9.89(-;1)  $&$-11.03$\\  
63& Eu      &		   &	          &  	         &		&	       &  	      &	             &$-9.89(-;1)  $&$-11.52$\\  
68& Er      &		   &	          &  	         &		&	       &  	      &	             &$-9.53(-;1)  $&$-11.11$\\  
70& Yb      &		   &	          &  	         &		&	       &  	      &	             &$-9.46(-;1)  $&$-10.96$\\  
\hline
&   \Teff   & 7400         & 7660	  & 7440         & 7328	        & 7138         & 7400	      & 7550         & 7800	    &	     \\
&   \logg   & 3.60         & 3.94	  & 4.08	 & 3.95	        & 3.92         & 4.00	      & 4.03         & 3.99	    &	     \\
&   \vmic   & 2.7          & 2.6	  & 2.7	         & 2.7	        & 2.8	       & 2.7	      & 2.7          & 3.6	    &	     \\
&   \vsini  & 179          & 163	  & 95	         & 200	        & 240	       & 100	      & 127          & 25	    &	     \\
\hline
\end{tabular}
}
\end{table*}


The line data required for our analysis were extracted from the VALD database. 
Line parameters of the lines selected for the fundamental parameter derivation 
and abundance analysis were checked by synthesising these lines in the solar 
spectrum taken from the National Solar Observatory 
Atlas\footnote{\tt http://www.coseti.org/natsolar.htm}. The lines for which we 
found a discrepancy between the observed and synthesised solar spectrum were 
rejected.

The typical number of lines selected for the analysis varied according to the 
rotational velocity of the analysed star. The Fe abundance was always 
determined on the basis of the largest number of lines, in comparison with 
other elements. The number of lines used to derive the abundance of each 
element is given in Table~\ref{abundance_table}, following the (internal) 
standard deviation associated with each measured abundance.

For all the elements that were not analysed we have adopted the solar abundance
from \citet{asplundetal2005}. (This fact enters into the model atmosphere 
computaion, and in the normalisation of the results expressed according to the 
total number of nuclei per unit volume $N_{\rm tot}$.)

\begin{figure}[ht]
\begin{center}
\includegraphics[width=90mm]{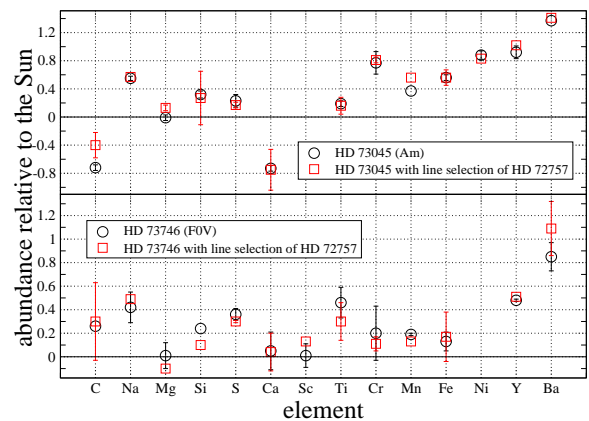}
\caption{Comparison between the abundances derived for 
HD~73045 (\vsini\, = 10 \kms -- open circles in the upper panel) and HD~73746 
(\vsini\, = 95 \kms\, -- open circles in the lower panel) adopting a proper 
line selection and the lines used to determine the abundances of 
HD~72757 (\vsini\, = 179 \kms\, -- open squares in both panels).} 
\label{lineselection} 
\end{center} 
\end{figure}

The high \vsini\, of some of the analysed stars resulted in a different and 
smaller selection of lines compared to the line list employed for slower 
rotators. We have checked if this is a source of systematic error by 
performing the abundance analysis of HD~73045 (\vsini\, = 10 \kms -- analysed 
in Paper~I) and HD~73746 (\vsini\, = 95 \kms) adopting the line list used for 
HD~72757 (\vsini\, = 179 \kms). The results of this test, illustrated in 
Fig.~\ref{lineselection}, show that no systematic errors are introduced by the 
difference in selected lines.

The high rotational velocity of many of the analysed stars results in a line
list dominated by strong and saturated lines. These lines are more sensitive to
\vmic\, variations than shallow lines. For this reason the uncertainties 
of the adopted \vmic, \Teff, and \logg\, values need to be considered 
to obtain a realistic estimate of the error bars associated with the derived 
element abundances. 

\begin{figure}[ht]
\begin{center}
\includegraphics[width=90mm]{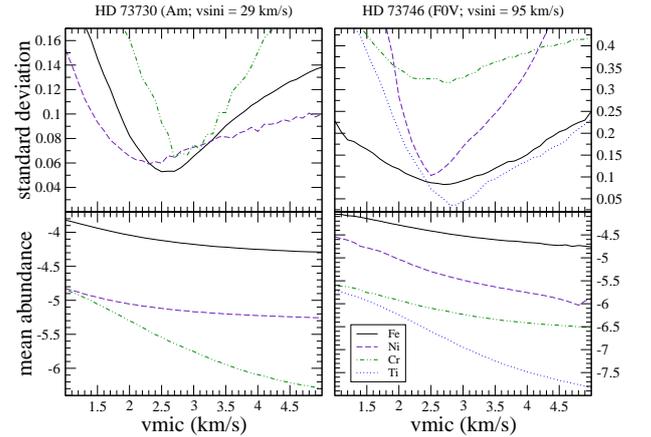}
\caption{The upper panels show the variation of the standard deviation from the
mean abundance of Fe, Ni, Cr and Ti as a function of \vmic, for HD~73730 
(\vsini\, = 29 \kms) and HD~73746 (\vsini\, = 95 \kms). The lower panels show
the variation of the mean abundance as a function of \vmic.} 
\label{vmic-slow rotators} 
\end{center} 
\end{figure}

\begin{figure}[ht]
\begin{center}
\includegraphics[width=90mm]{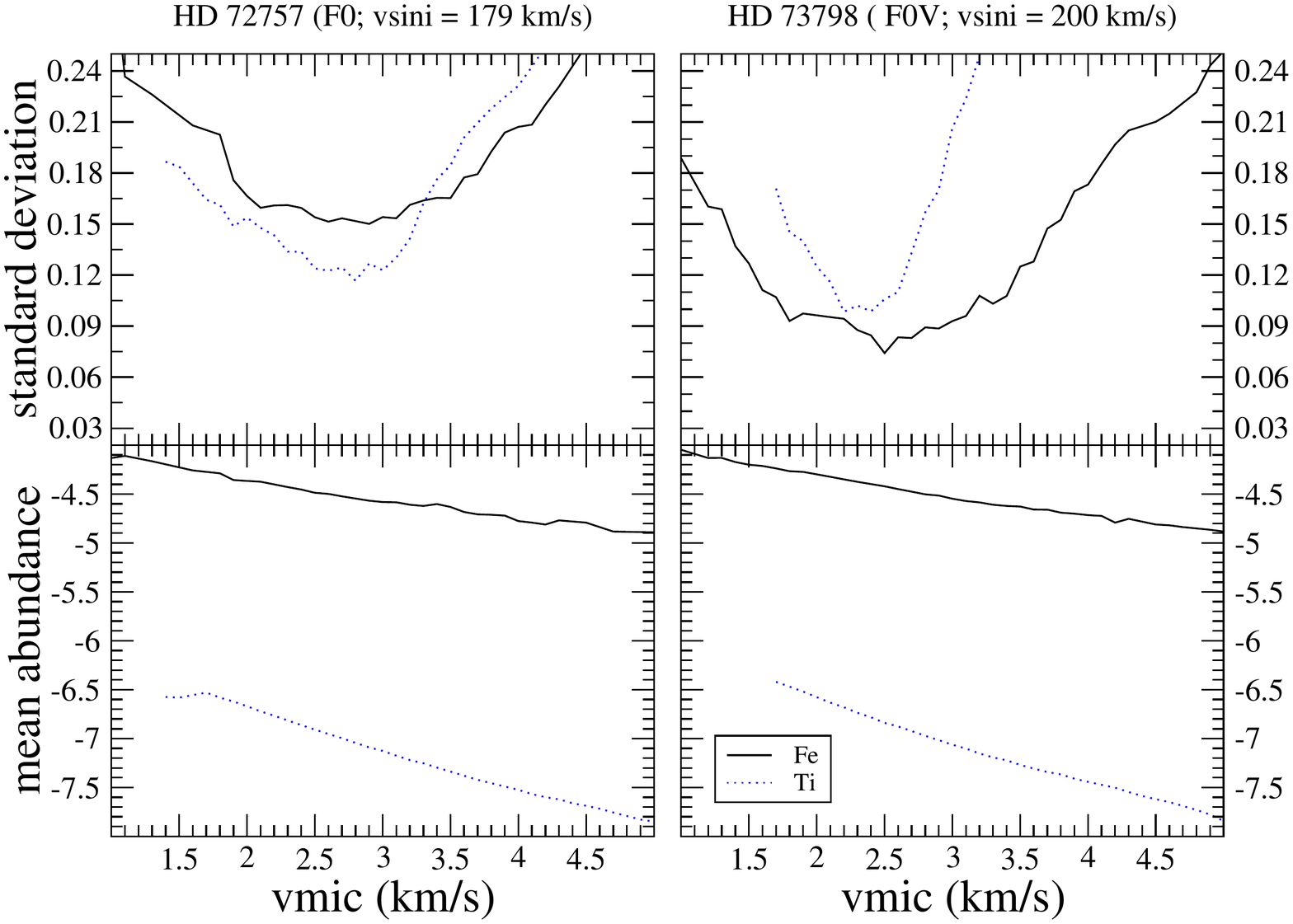}
\caption{As for Fig.~\ref{vmic-slow rotators}, but for HD~72757 
(\vsini\, = 179 \kms) and HD~73798 (\vsini\, = 200 \kms).} 
\label{vmic-fast rotators} 
\end{center} 
\end{figure}

In Paper~I we have 
shown the results of two tests performed on slowly rotating stars. They show 
i) how the standard deviation from the mean Fe abundance changes for 
\vmic\, = 0, 1, 2, ..., 6 \kms, adopting fundamental parameters derived from
photometry and spectroscopy and ii) the uncertainties introduced in the
abundances of Fe, Ti and Ni by varying effective temperature and 
\logg\, by $\pm$200~K and $\pm$0.2~dex respectively. Our tests show that the 
spectroscopic parameters are comparable or better than the photometric 
parameters. Furthermore, the tests show a variation of less than 0.2~dex in 
abundance due to the temperature variation. No significant abundance change was 
found varying when \logg. However, the \vmic\, value and its associated 
uncertainty increase their importance for rapidly rotating stars, and for this 
reason we have performed the following additional test. For four stars with 
different \vsini\, values (HD~73730, \vsini\, = 29 \kms -- HD~73746, 
\vsini\, = 95 \kms -- HD~72757, \vsini\, = 179 \kms -- HD~73798, 
\vsini\, = 200 \kms) we have performed the abundance analysis of selected Fe, 
Ni, Cr and Ti lines for assumed \vmic\, varying from 1 to 5 \kms\, in steps 
of 0.1 \kms. We have then plotted in Fig.~\ref{vmic-slow rotators} and 
in Fig.~\ref{vmic-fast rotators} the variation of the standard deviation from 
the mean abundance (upper panel) and the mean abundance itself (lower panel) 
as a function of \vmic.

Figures \ref{vmic-slow rotators} and \ref{vmic-fast rotators} show that the 
adopted \vmic\, value (2.7 \kms\, for all the stars taken
into account in this test) obtained with Eq.~\ref{vmicphotometry} is 
reasonable at every rotational velocity. The only exception is for the Am 
stars, for which slightly larger values of \vmic\, are obtained for those 
stars for which this quantity is directly derived. Taking into account the 
different best \vmic\, values obtained from different elements, and the 
dispersion of the standard deviation as a function of \vmic, the 
uncertainty that we assign to \vmic, for the rapid rotators, is 0.7 \kms. 
The lower panels of Fig.~\ref{vmic-slow rotators} and 
Fig.~\ref{vmic-fast rotators} allow us to derive directly the abundance 
uncertainty given the uncertainty of \vmic\, for each element. We have 
used Fe as a standard element to derive the abundance standard error, since it 
is the element with the highest number of selected lines. An error bar of 
0.7 \kms\, in \vmic\, produces an uncertainty of element abundance increasing 
from 0.1 dex for the slow rotators, to 0.2 dex for the rapid rotators. The 
error bar on the element abundance is dominated by the uncertainty of 
\Teff\, for the slow rotators, while the uncertainty in \vmic\, becomes 
increasingly important for larger \vsini. Assuming that \Teff\, and 
\vmic\, are completely uncorrelated, the resultant abundance error bar
for the rapidly rotating stars is given by standard error propagation theory,
which leads to an uncertainty of about 0.3 dex. 
In conformity with this analysis, we assume an abundance error bar 
increasing linearly with \vsini\, from 0.2 dex to 0.3 dex at a \vsini\, value 
of 200 \kms.  

Another source of uncertainty that is dependent on \vsini\, is due to the 
continuum normalisation. To quantify this uncertainty we have performed the 
following test. We have derived the abundance of the \ion{Fe}{ii} line at 
5325.6~\AA\, for HD~73746 (\vsini\, = 95 \kms), HD~72757 (\vsini\, = 179 \kms) 
and HD~73798 (\vsini\, = 200 \kms) with the adopted normalised observed 
spectrum and with the spectrum multiplied/divided by 0.99. In this way we have 
increased/decreased the continuum level of 1\%, that we estimate to be a 
reasonable uncertainty. The difference between the abundances obtained in this 
way is increasing with \vsini\, from about 0.1~dex for HD~73746, to about 
0.2~dex for the two faster rotators. We believe that in most cases this error 
bar is an upper limit since the line by line abundance analysis method, used 
in this work, allows a very careful line selection that rejects all the lines 
for which the continuum level looks uncertain. Under the assumption of no 
systematic errors in the continuum normalisation, this uncertainty is 
decreasing with an increase of the number of selected lines. If for some lines 
the continuum level is too low, for some others it would be too high, leading 
to an increase of the dispersion, but not to an abundance variation. 

\begin{figure*}[ht]
\begin{center}
\rotatebox{270}{
\includegraphics[width=230mm]{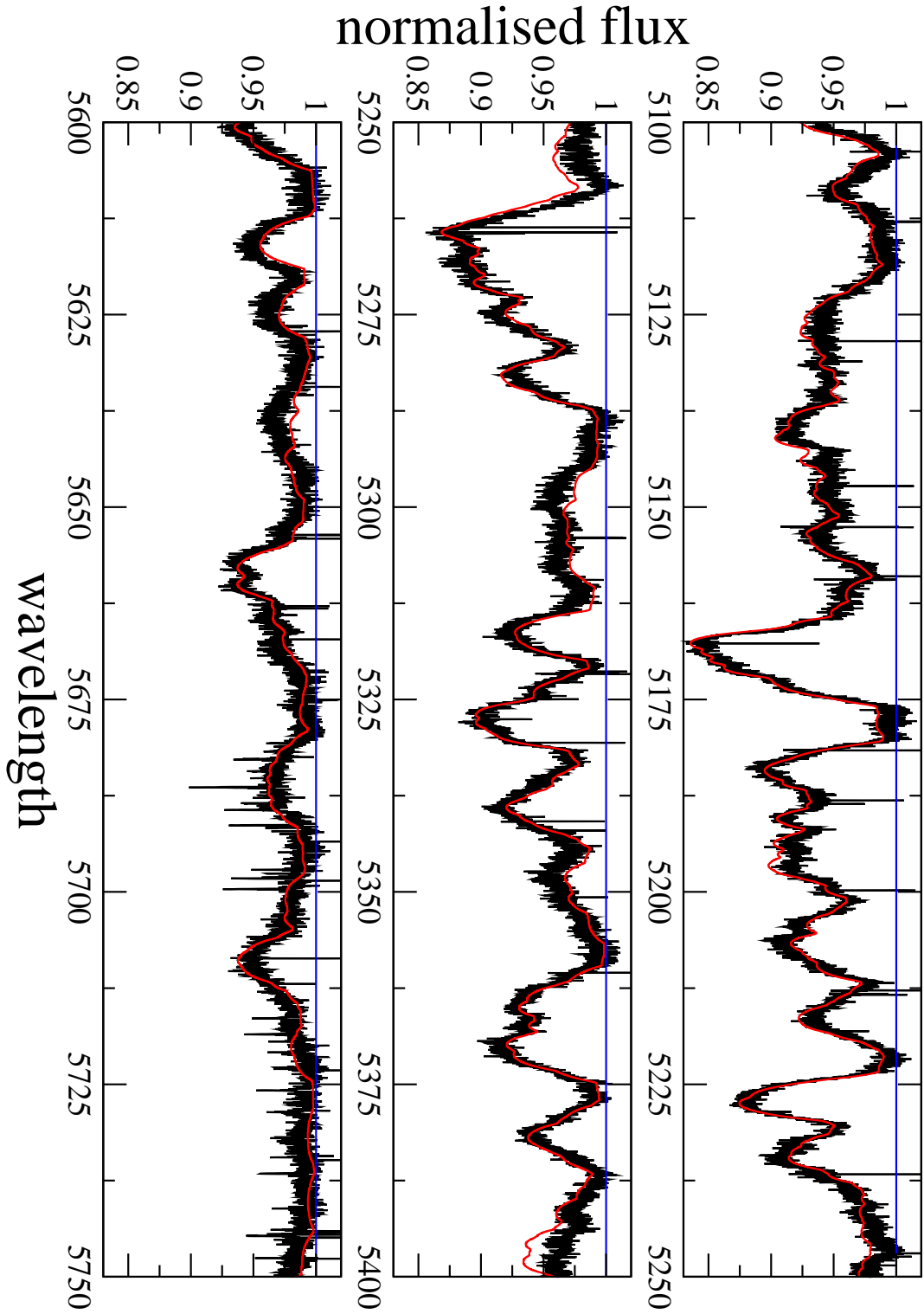}}
\caption{Portion of the spectrum of HD~72757 (\vsini\, = 179 \kms) and the
synthetic spectrum after the complete abundance analysis in black and red
respectively.} 
\label{spettro} 
\end{center} 
\end{figure*}

An example of the spectrum of a rapidly rotating star and the synthetic 
spectrum obtained after the complete abundance analysis is shown in 
Fig.~\ref{spettro} (available online).

For all the rapid rotators in our sample, we were able to obtain only an upper 
limit on the abundance of Li, because the high \vsini\, values did not allow 
us to detect the Li line. Only the Li abundance of HD~74656 may be considered 
as a real estimate, although even this value is uncertain because it was 
derived only from one blended line.

We confirm the membership in Praesepe of all the stars of our sample, because 
the \vr\, measurements are all compatible with the cluster mean of 
34.5$\pm$0.0 \kms\, \citep{robichon}. For none of the analysed stars have we 
found any evidence of binarity. These considerations confirm the quality of our 
target selection (which were selected for membership and
non--binarity).

\begin{figure}[ht]
\begin{center}
\includegraphics[width=90mm]{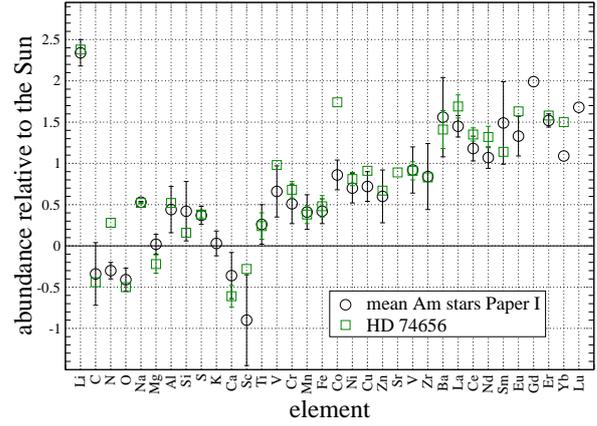}
\caption{Comparison between the mean abundance of the Am stars analysed in 
Paper~I and the Am star HD~74656. This plot confirms the Am classification 
of this star.} 
\label{compare Am-HD74656} 
\end{center} 
\end{figure}

We confirm the Am classification of HD~74656. In Fig.~\ref{compare Am-HD74656}
 (available online) the mean abundances for the Am stars analysed in Paper~I 
are compared with the abundances obtained for HD~74656. The comparison shows a 
good agreement between the two abundance patterns. 
\section{Discussion}\label{discussion}
In this section we discuss the results obtained considering both samples of 
stars: those analysed in this work, and those from Paper~I. In the sample of 
stars from Paper~I, we have not included results for HD~73666 (Blue
Straggler), HD~72942 (probable non--member) and HD~73174, for which the 
derived \Teff\, from Str\"omgren photometry and from spectroscopy differ by 
about 700~K and no explanation was found in Paper~I. For this reason we have 
decided not to consider this object in our final analysis of the cluster.
\subsection{Abundances of the A$-$ and F$-$type stars of the Praesepe
cluster}\label{Abundances of the A- and F-type stars}
\begin{figure}[ht]
\begin{center}
\includegraphics[width=90mm]{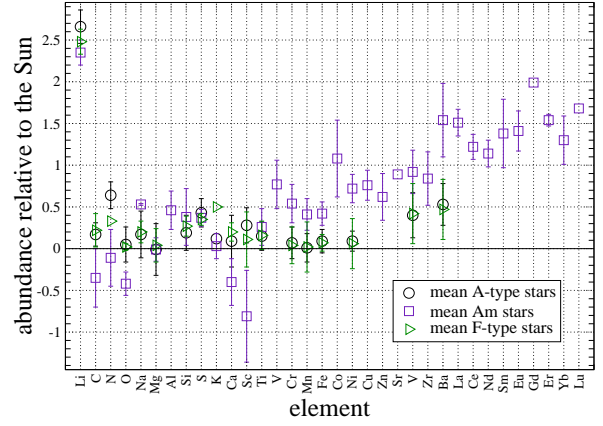}
\caption{Comparison between the mean abundances obtained for Am, A-- and
F--type stars indicated by an open circle, square and triangle respectively. 
The error bars are the standard deviations from the calculated mean abundances. 
In the case of only one measurement, no error bar is shown. A-- and F--type 
stars show a similar abundance pattern. The Am stars show the typical peculiar 
abundance pattern described in detail in Paper~I.} 
\label{mean.abundances} 
\end{center} 
\end{figure}

The stars were first grouped according to their classifications as F, A and Am 
stars; Fig.~\ref{mean.abundances} shows the mean abundances for each group. 
The error bars in Fig.~\ref{mean.abundances} are the standard deviations 
from the calculated mean abundances of each group.

The A-- and F--type stars show similar abundance patterns, characterised
by solar abundances for almost all the elements. The Am stars show the typical 
abundance pattern characterised by underabundances of C, N, O, Ca and Sc, with 
a general overabundance of the other elements. As expected, the error bars 
associated with the abundances of the F--type stars turn out to be comparable 
to or smaller than those for the A--type stars. It is also possible to see 
this effect in Figures from \ref{trend1} to \ref{trend2}.

We have calculated the metallicity of the cluster (Z) from the abundances 
derived for the F--type stars, excluding HD~73575. This star is close to the 
TAMS and its Fe abundance differs by 0.27 dex with respect to the mean Fe 
abundance (0.11$\pm$0.03 dex) of the other F--type stars. For this reason we 
have omitted HD~73575 from the determination of the cluster metallicity. The Z 
value we obtain is 0.015$\pm$0.002 dex.

The Z value adopted by several other authors and used to characteris isochrones
(see Sect.~\ref{HRdiagram}) is calculated with the following approximation:
\begin{equation}
\label{clusterZ}
Z_{\mathrm cluster} \simeq 10^{([FE/H]_{\mathrm Fstars}-[Fe/H]_{\odot})} \cdot Z_{\odot}\ ,
\end{equation}
assuming $Z_{\odot}$=0.019 dex. We have recalculated the Z of the cluster
according to this approximation obtaining Z = 0.024$\pm$0.02. We have 
derived this value using the mean Fe abundance of all the F--type stars except 
HD~73575 ([Fe/H]=0.11$\pm$0.03 dex). 
 
Our estimated metallicity is in agreement with that included in the catalogue 
of \citet{chen2003} of [Fe/H]=0.14 dex and with that recently obtained by 
\citet{deokkeun} of [Fe/H]=0.11$\pm$0.03 dex. \citet{deokkeun} derived the
metallicity from an Fe abundance determination from high resolution spectra of 
four G--type cluster members.
\subsection{HR diagram}\label{HRdiagram}
\begin{figure}[ht]
\begin{center}
\includegraphics[width=90mm]{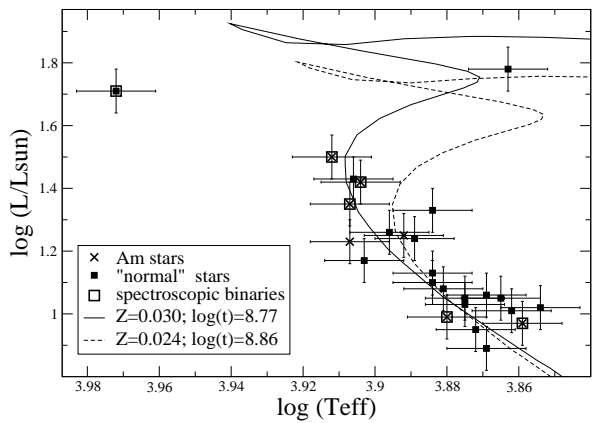}
\caption{Hertzsprung--Russell diagram of the Praesepe cluster. The crosses 
and filled squares show the position on the HR diagram for the Am and normal 
A-- and F--type stars, respectively. The open squares indicate the 
spectroscopic binaries. We have not corrected the luminosity of these stars 
due to the presence of a companion, but we assume it less than 0.1 dex. The 
error bar in luminosity is 0.07 dex. The dashed line shows an isochrone from 
\citet{girardi2002} for the age and metallicity given in the literature 
(\logt\, = 8.85 dex; Z = 0.024 dex). The solid line shows the isochrone 
corresponding to our best fit (\logt\, = 8.77 dex; Z = 0.030 dex).} 
\label{hr_diagram} 
\end{center} 
\end{figure}

Using the spectroscopic temperatures determined to perform the abundance
analysis we have built the Hertzsprung--Russell (HR) diagram of the cluster 
around the turn--off point (Fig.~\ref{hr_diagram}).
We have calculated the luminosity of each star photometrically, adopting the
magnitudes in $V$ band given by SIMBAD and a cluster distance modulus of 
6.30$\pm$0.07 mag \citep{vonLeeuwen}. We have used a reddening of 0.009 mag 
(WEBDA) and the bolometric correction given by \citet{balona}. Luminosities 
are listed in Table~\ref{table luminosity}. With a distance modulus 
uncertainty of 0.07 mag, an uncertainty in the bolometric correction of about 
0.07 mag, and a reddening uncertainty of 0.01 mag, we estimate that the 
typical uncertainty in \mbol\, is about 0.15 mag, corresponding to an 
uncertainty in \logl\, of about 0.06 dex.

In the diagram we have divided our sample into Am stars (crosses) and normal
A-- and F--type stars (full squares). We have also denoted (open squares) the
spectroscopic binary stars. We have not corrected the luminosities of the
spectroscopic binaries, as the only information we have about the secondaries
is that their contribution to the analysed spectra is not detected (Paper~I). 
This suggests that their flux contribution is below about 5\%. 

The two stars located at the top left and top right of the HR diagram are 
HD~73666 and HD~73575, respectively. Both have already been analysed in detail 
in Paper~I. HD~73666 is clearly a Blue Straggler, while HD~73575 should be 
just at the TAMS, if the stars are cluster members (membership given by 
HIPPARCOS and confirmed in Paper~I). HD~73575 deserves a further comment. It 
is the most evolved star of the cluster and the position close to the TAMS is 
consistent with this. This star also shows anomalous chemical composition 
compared with that of the other normal F--type stars. The position on the HR 
diagram and the peculiar chemical composition of HD~73575 suggest an unusual 
evolutionary history.

We have adopted isochrones by \citet{girardi2002} to determine age and
metallicity of the cluster using the HR diagram. In Fig.~\ref{hr_diagram} we
have plotted two isochrones: one correspoding to the age and metallicity of 
the cluster taken from WEBDA (\logt\, = 8.85 dex; Z = 0.024 dex; dashed line) 
and the other corresponding to our best fit on the HR diagram 
(\logt\, = 8.77 dex; Z = 0.030 dex; solid line). 

\citet{deokkeun} determined the metallicity of the Praesepe cluster by fitting 
Yale Rotating Evolutionary Code \citep[YREC,][]{sills2000} isochrones to 
the main sequence of the cluster, obtained from Johnson photometry. They 
derived a metallicity of [Fe/H] = 0.20$\pm$0.04 corresponding to 
Z = 0.030$\pm$0.003. This result is in agreement with the results of our fits 
of isochrones to our sample of stars around the turn--off point.

All the Am stars are on the main sequence. Only HD~73618 (the hottest
main sequence star in Fig.~\ref{hr_diagram}) can be considered a Blue
Straggler \citep{BS2007}, taking into account the age and metallicity given 
in the literature. 
\subsection{Abundances vs. \Teff\, and \vsini\,}\label{correlation parameters}
\begin{figure*}[ht]
\begin{center}
\includegraphics[scale=0.7]{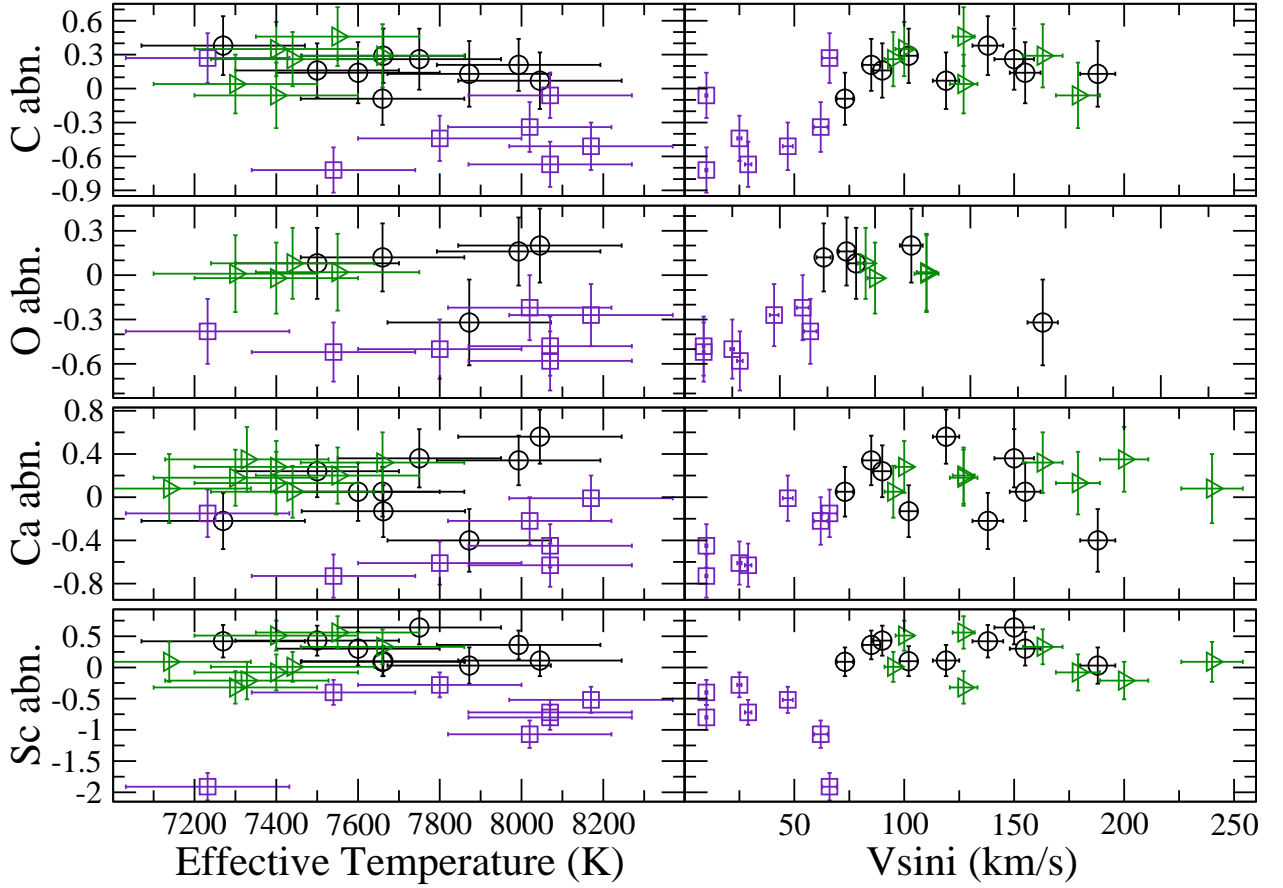}
\caption{Abundances relative to the Sun \citep{asplundetal2005} of C, O, Ca and
Sc as a function of \Teff\, and \vsini\, for normal A--type stars (circle), Am 
stars (square) and normal F--type stars (triangle). A correlation between 
abundances of C, O, Ca and \vsini\, is found for Am stars; an anticorrelation 
between Sc abundance and \vsini\, is found for Am stars.} 
\label{trend1} 
\end{center} 
\end{figure*}

\begin{figure*}[ht]
\begin{center}
\includegraphics[scale=0.7]{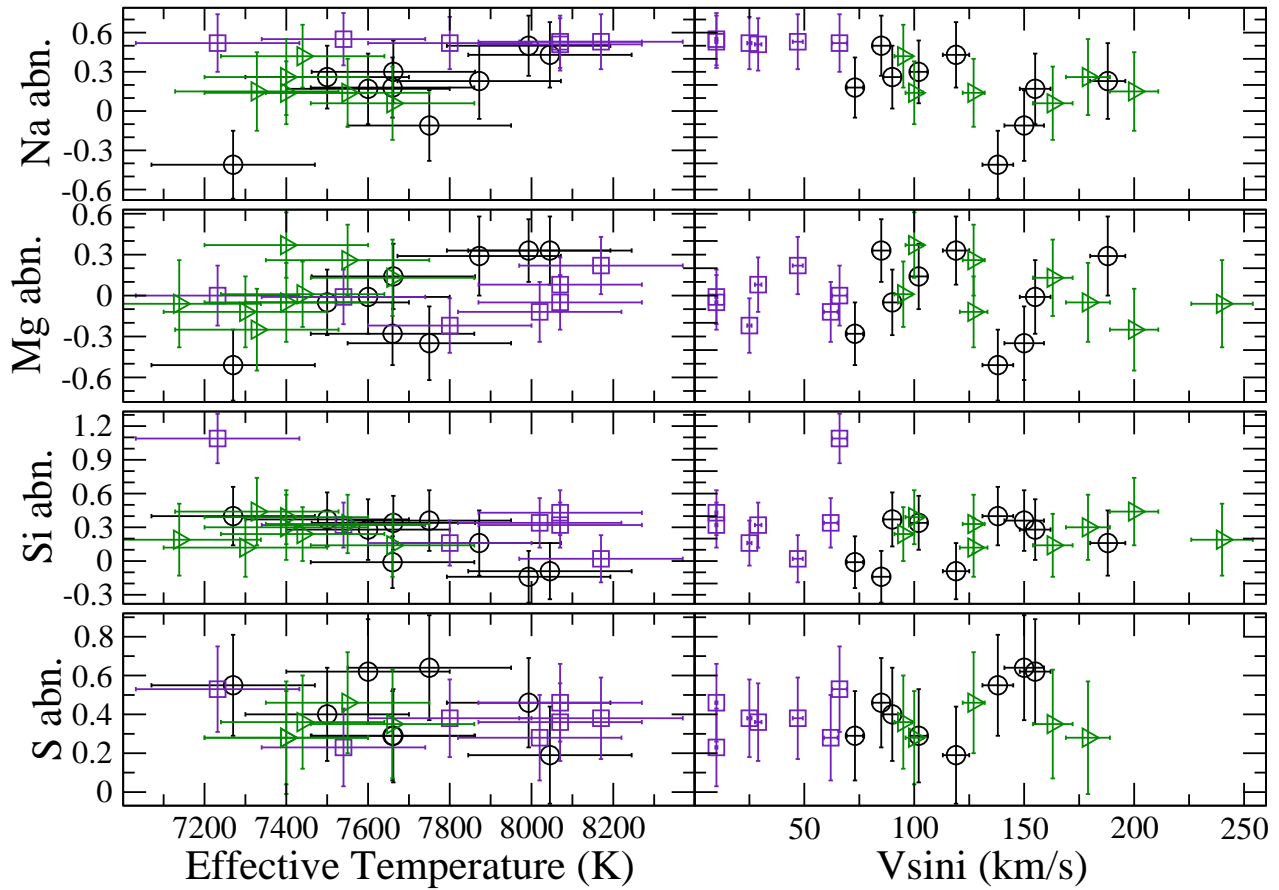}
\caption{Same as in Fig.~\ref{trend1} but for Na, Mg, Si and S. No correlation 
is found.} 
\label{notrend} 
\end{center} 
\end{figure*}

\begin{figure*}[ht]
\begin{center}
\includegraphics[scale=0.7]{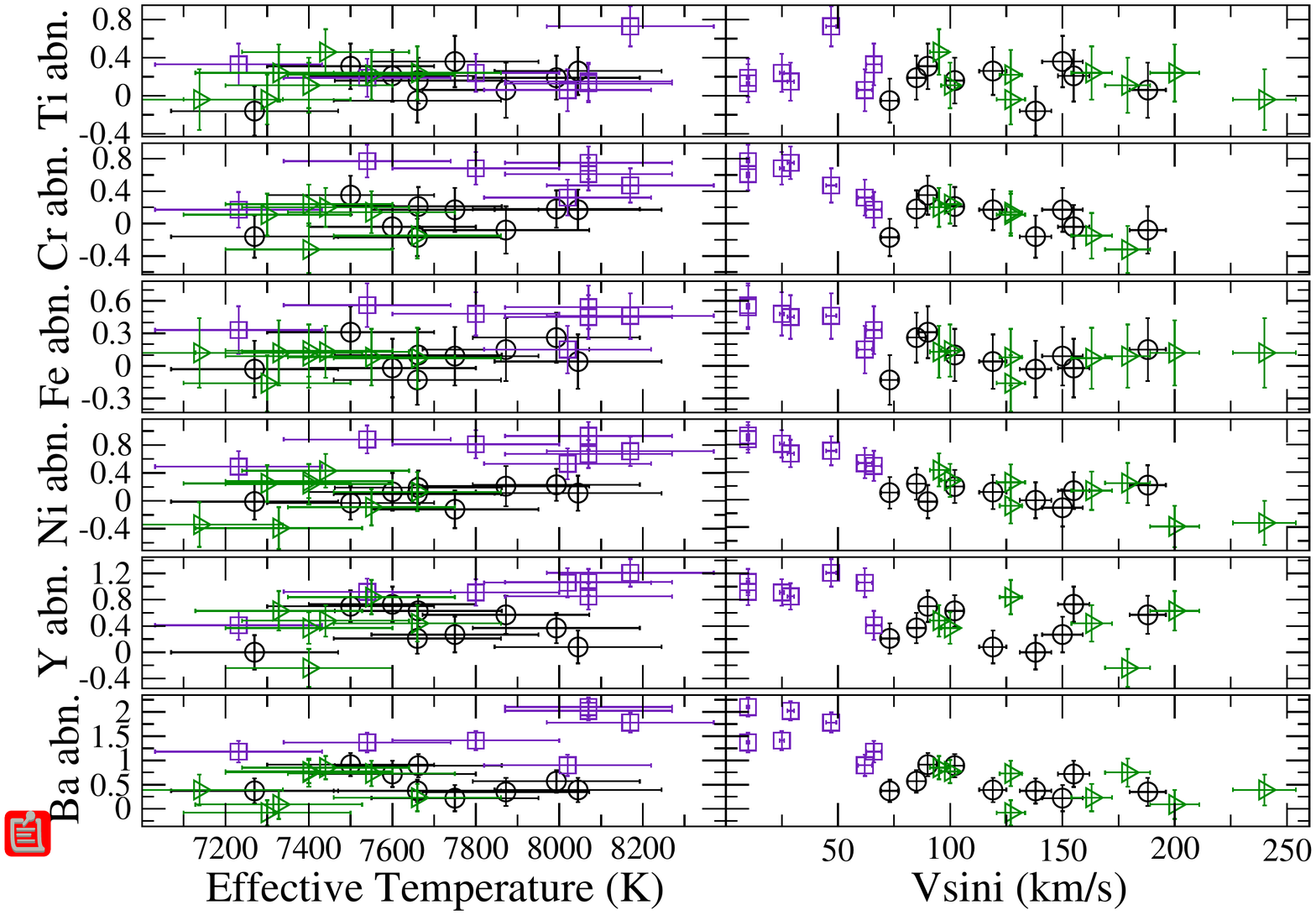}
\caption{Same as in Fig.~\ref{trend1} but for Ti, Cr, Fe, Ni, Y and Ba. No 
correlation is found for Ti. An anticorrelation between Cr, Fe, Ni, Y and Ba 
abundance and \vsini\, is found for Am stars.} 
\label{trend2} 
\end{center} 
\end{figure*}

The abundances of the elements analysed for most of the stars are displayed 
against \Teff\, and \vsini\, in the left and right panels respectively from 
Fig.~\ref{trend1} to Fig.~\ref{trend2}. We have divided the sample of stars 
according to their spectral classification: A--type stars (open circles), Am 
stars (open squares) and F--type stars (open triangles). This division and 
visualisation will be retained in the following sections.

The plots show that the abundances of the F--type stars are less 
scattered than those obtained for the A--type stars, as expected. 

For the normal A-- and F--type stars we do not find any clear correlation 
between abundance and \Teff\, or abundance and \vsini. This result confirms 
the prediction of \citet{charbonneau}. 

For the Am stars, none of the plotted elements, except Y, shows any trend 
between abundance and \Teff. \citet{richer2000} predict mild correlations 
between abundances of some elements (e.g. Ni) and \Teff, but probably our 
temperature range is not large enough to show such correlations.

In Am stars, we find a strong correlation between abundance and
\vsini\, for all peculiar elements, except for
scandium and titanium. As described in detail in Paper~I, Am stars of the
Praesepe cluster show peculiarities mainly for C, O, Ca and Sc (underabundant)
and Ti, Cr, Fe, Ni, Y and Ba (overabundant). C, O and Ca show an increasing
abundance with \vsini. The Fe--peak elements, Y, and Ba show a decreasing
abundance with \vsini. Na, Mg, Si and S are not peculiar in Am stars and they
do not show any correlation with \vsini. As a general behavior, the
peculiarities decrease with \vsini, becoming closer and closer to the
abundances typical of the normal A-- and F--type stars of the cluster.
Scandium and titanium are exceptions: Ti does not show any clear trend with 
\vsini, and Sc shows a trend opposite to that of C, O and Ca, while we 
would have expected a similar behavior.

\citet{burkhart1979} derived the abundance anomaly, from the Str\"omgren 
m1 index, and \vsini\, values for a sample of well-known Am stars. She
suggested a jump in the abundance anomaly for stars with \vsini\, $>$ 55 \kms;
our results confirm this claim.

Models by \citet{charbonneau} do not predict any strong correlation between 
element abundances and rotational velocity: "For FmAm stars, it was shown 
quantitatively that despite the potentially inhibiting effects of meridional 
circulation on chemical separation, no dependence of abundance anomalies on 
\vsini\, is expected, for most elements, in the rotational velocity interval 
characteristic of FmAm stars.". The correlations here obtained between 
abundance and \vsini, for the Am stars, are in clear disagreement with the 
predictions of \citet{charbonneau}. 

\citet{richer2000} found that element abundances in model Am stars are 
dependent only on the depth of the zone mixed by turbulence, thus showing that 
the abundance anomalies in Am stars depend mainly on the turbulence model. 

More recently, \citet{talon} tested rotational mixing models with both the 
Geneva--Toulouse and Montreal codes. They showed how the predicted surface 
abundances of FmAm stars should vary as a function of rotational velocity 
of the star. Our findings support their predictions.
\subsection{Abundances vs. M/M$_\odot$ and fractional age}
\label{correlation mass and fractional age}
For the stars in our sample we have calculated M/M$_{\odot}$, fractional age 
($\tau$) and their uncertainties according to \citet{john2006} and listed them 
in Table~\ref{table luminosity}. 
\begin{table}[ht]
\caption[]{\logl, $\log$\Teff, M/M$_{\odot}$ and fractional age ($\tau$) with associated error bars 
for the analysed stars of the Praesepe cluster. The error bar in luminosity is 
$\sim$0.06 dex, as explained in Sect.~\ref{HRdiagram}. No fractional age is given for HD~73666 since
it is a Blue Straggler.}
\label{table luminosity}
\scriptsize{
\centering                      
\begin{tabular}{ccccccccc}
\hline
\hline
HD & \logl & Sp.Type & $\log$\Teff & M/M$_{\odot}$ & $\sigma_{M/M_{\odot}}$ & $\tau$ & $\sigma_\tau$ \\
   &   dex &         &             &               &                        &        &               \\
\hline
72757 & 1.06 & F0     & 3.869 & 1.71 & 0.09 & 0.43 & 0.06 \\
72846 & 1.43 & A5V    & 3.906 & 2.09 & 0.15 & 0.76 & 0.11 \\
73045 & 0.99 & Am     & 3.880 & 1.67 & 0.08 & 0.40 & 0.06 \\
73175 & 1.13 & F0V    & 3.884 & 1.78 & 0.09 & 0.48 & 0.07 \\
73345 & 1.17 & A7V    & 3.903 & 1.84 & 0.13 & 0.53 & 0.08 \\
73430 & 1.10 & A7V    & 3.884 & 1.76 & 0.09 & 0.47 & 0.07 \\
73450 & 1.01 & A9V    & 3.862 & 1.66 & 0.08 & 0.39 & 0.06 \\
73574 & 1.33 & A8V    & 3.884 & 1.96 & 0.10 & 0.64 & 0.10 \\
73575 & 1.78 & F0III  & 3.863 & 2.33 & 0.12 & 1.02 & 0.15 \\
73618 & 1.50 & Am     & 3.912 & 2.16 & 0.22 & 0.84 & 0.13 \\
73666 & 1.71 & A0     & 3.972 & 2.46 & 0.12 &      & 0.18 \\
73709 & 1.35 & Am     & 3.907 & 2.00 & 0.14 & 0.68 & 0.10 \\
73711 & 1.42 & Am     & 3.904 & 2.08 & 0.10 & 0.75 & 0.11 \\
73730 & 1.23 & Am     & 3.907 & 1.89 & 0.13 & 0.58 & 0.09 \\
73746 & 0.95 & F0V    & 3.872 & 1.64 & 0.08 & 0.38 & 0.06 \\
73798 & 1.05 & F0V    & 3.865 & 1.70 & 0.12 & 0.42 & 0.06 \\
73818 & 0.97 & Am     & 3.859 & 1.63 & 0.08 & 0.37 & 0.06 \\
73993 & 1.02 & F2V    & 3.854 & 1.66 & 0.12 & 0.39 & 0.06 \\
74028 & 1.24 & A7V    & 3.889 & 1.88 & 0.09 & 0.57 & 0.09 \\
74050 & 1.26 & A6V    & 3.896 & 1.91 & 0.10 & 0.60 & 0.09 \\
74135 & 0.89 & F0     & 3.869 & 1.60 & 0.08 & 0.35 & 0.05 \\
74587 & 1.03 & A5     & 3.875 & 1.71 & 0.09 & 0.43 & 0.06 \\
74589 & 1.05 & F0     & 3.875 & 1.70 & 0.09 & 0.42 & 0.06 \\
74656 & 1.25 & Am     & 3.892 & 1.89 & 0.09 & 0.58 & 0.09 \\
74718 & 1.08 & A5     & 3.881 & 1.74 & 0.09 & 0.45 & 0.07 \\
\hline
\end{tabular}
}
\end{table}



\begin{figure}[ht]
\begin{center}
\includegraphics[width=95mm]{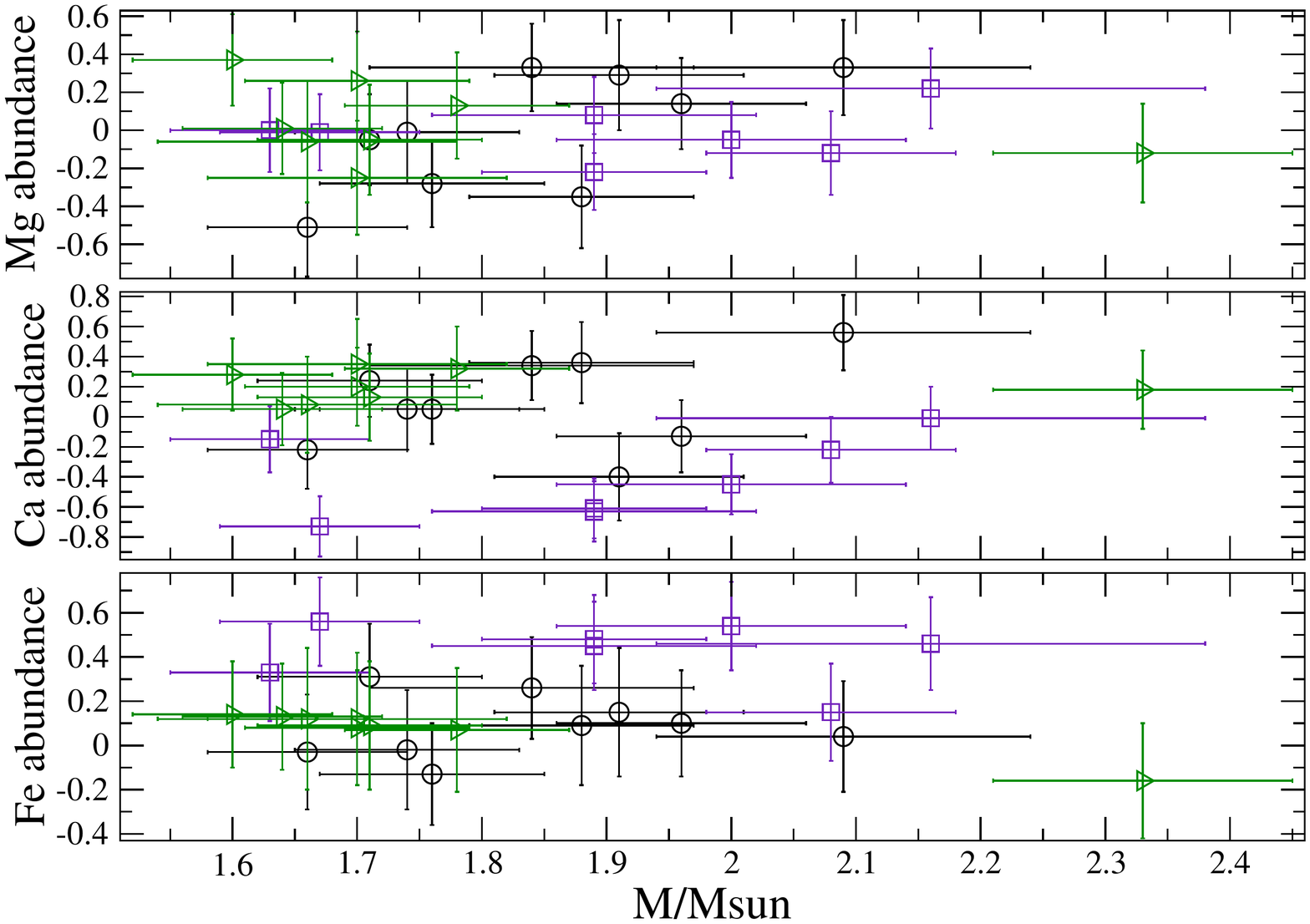}
\caption{Abundances relative to the Sun \citep{asplundetal2005} of Mg, Ca and 
Fe as a function of the M/M$\odot$. No correlations are found.} 
\label{CONaMgSSivsMass} 
\end{center} 
\end{figure}

We have taken the abundances of three representative elements (Mg, Ca and Fe) 
showing, for the Am stars, the three different properties discussed in 
Sect.~\ref{correlation parameters}. We plot the abundances of these elements 
as functions of M/M$_{\odot}$ and display the results in 
Fig.~\ref{CONaMgSSivsMass}. 

No clear correlation is visible for any of the three plotted elements, for the 
normal or the Am stars. 

\citet{richer2000} suggest that the abundance anomalies typical of Am stars
depend very little on the mass and therefore fractional age of the star. 
This is consistent with the results of this study.
\subsection{Searching for correlations between abundances}
\label{correlation abundances}
In a sample of seventeen Am stars, \citet{adelman2007} found significant 
correlations between the abundances of twelve elements. 

\begin{figure}[ht]
\begin{center}
\includegraphics[width=95mm]{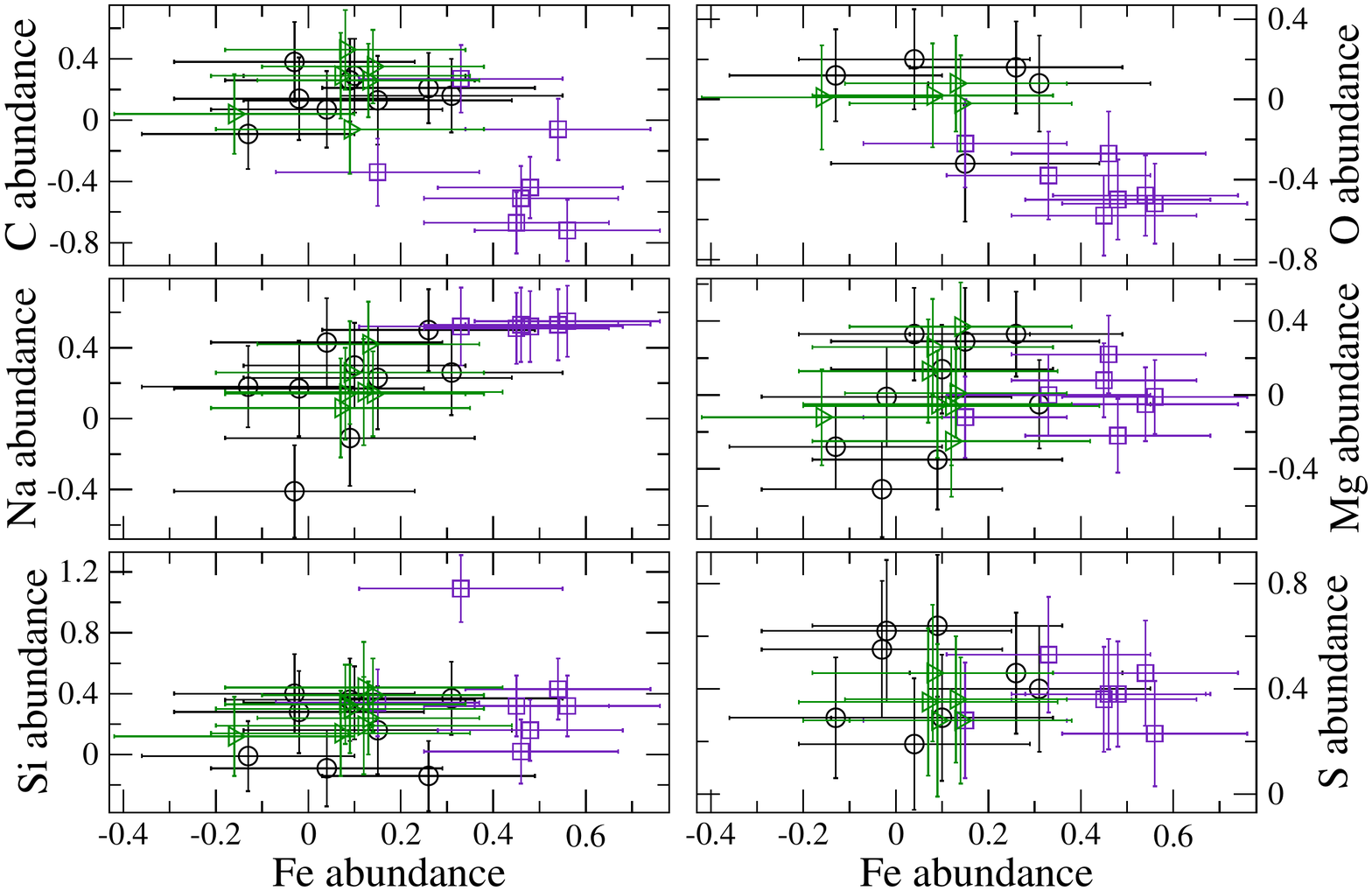}
\caption{Abundances relative to the Sun \citep{asplundetal2005} of C, O, Na, 
Mg, S and Si as a function of the Fe abundance.} 
\label{CONaMgSSivsFe} 
\end{center} 
\end{figure}

\begin{figure}[ht]
\begin{center}
\includegraphics[width=95mm]{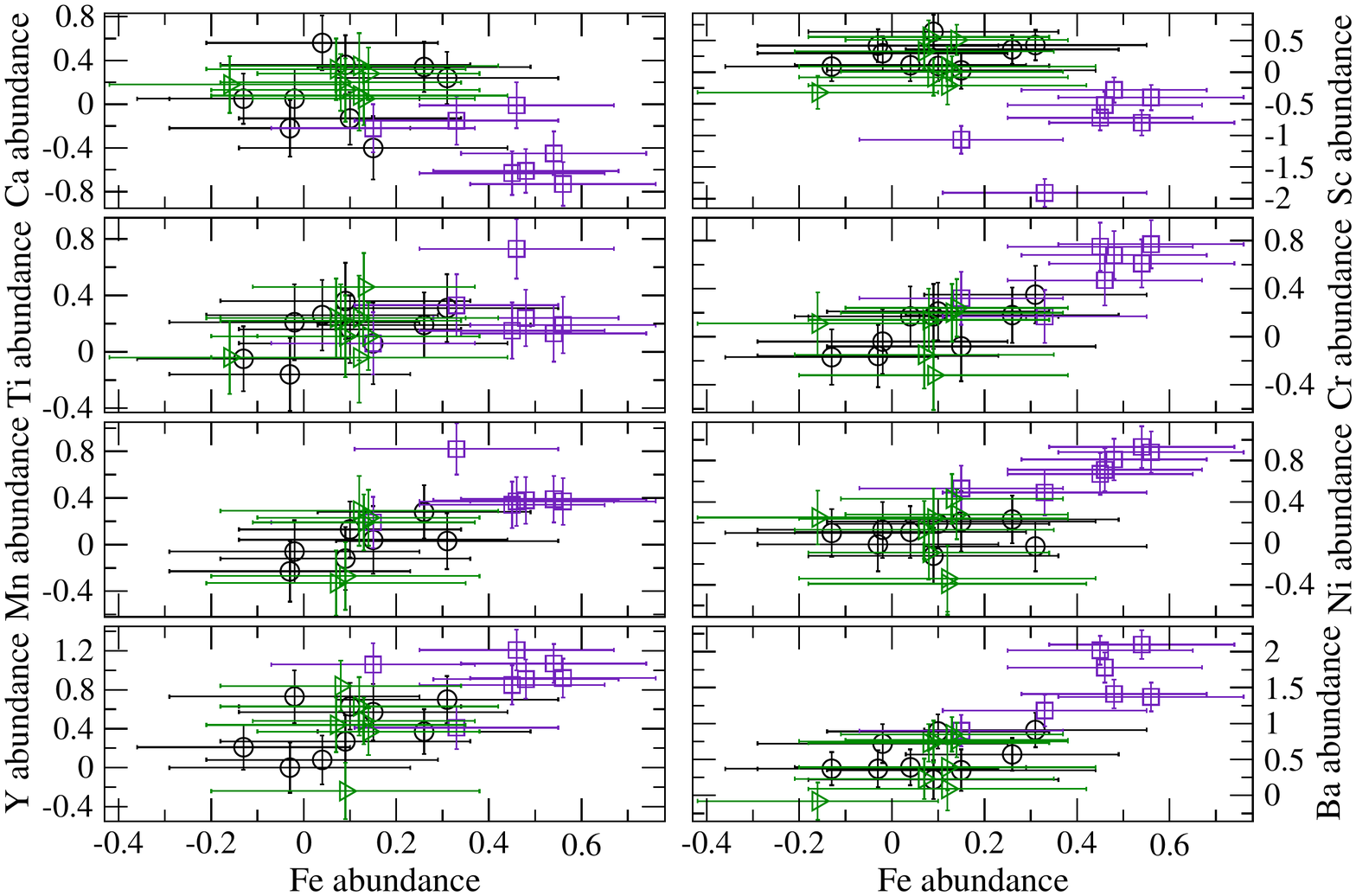}
\caption{Same as in Fig.~\ref{CONaMgSSivsFe} but for Ca, Sc, Ti, Cr, Mn, Ni,
Y and Ba.} 
\label{CaScTiCrMnNiYBavsFe} 
\end{center} 
\end{figure}

The abundances of the elements analysed in most of the stars of our sample are
displayed against the Fe abundance in Fig.~\ref{CONaMgSSivsFe} and 
Fig.~\ref{CaScTiCrMnNiYBavsFe}.

We do not find any clear correlation between the abundances of any element and
Fe for the A-- and F--type stars due to the tiny spread in Fe abundance.

In contrast, the Am stars show correlations for all the elements in which a 
correlation was already found with \vsini. The Fe abundance is an indicator 
of Am peculiarities. As we have found a correlation between \vsini\, and Am 
peculiarities, it is not surprising to obtain similar correlations with the 
Fe abundance.

The correlations found for C and O are in agreement with those found by 
\citet{igor} in a sample of seventeen well--known Am stars analysed by 
different authors. They looked for a correlation between the C abundance 
and \vsini, but without any positive result.

From our results it is clear that these correlations are related to those
found with \vsini\, in Sect.~\ref{correlation parameters}.
\section{Conclusions}\label{conclusions}
We have obtained high resolution, high SNR spectra for seventeen F-- and 
A--type stars of the Praesepe open cluster, with the SOPHIE spectrograph at OHP.
One of these stars (HD~74656) was classified as Am star.

We have calculated the fundamental parameters and performed a detailed 
abundance analysis for all the stars of the sample, confirming the previous
classification, as chemically peculiar, of the Am star HD~74656.

The stars we observed were selected using a standard procedure that
will be adopted for future analysis of other open clusters. All the 
selected stars of the sample shown in Table ~\ref{tabella radec} are 
confirmed members of the cluster and among them there are no known binaries 
(except HD~73666). This shows the quality of the selection procedure, since 
membership and non--binarity were required for the selection.

To be able to discuss general properties of the cluster we have added to our
sample of stars those considered in Paper~I. Of the stars analysed in 
Paper~I we have not included HD~73666 (Blue Straggler), HD~72942 (not a 
confirmed member) and HD~73174 (important uncertainties in \Teff). The
sample of stars analysed in Paper~I and in this work contains all the known 
cluster members with a spectral type between F2 and A1 and not SB2. This 
removes any bias that could have been introduced by an inappropriate target 
selection.

We have divided our final sample of stars according to their spectral
classification, taken from WEBDA, and derived the mean abundance pattern for 
three groups of stars: A--, F--type and Am. The A-- and F--type stars show a
similar abundance pattern, close to solar, while the Am stars show the typical 
Am peculiarities described in Paper~I. The abundance scatter of the Am stars 
is higher than for the normal stars and, between A-- and F--type stars, the 
F--stars show the smaller scatter, as expected.

We have derived the metallicity of the cluster from the F--type stars of our
sample, omitting HD~73575 which is already on the TAMS. The
resultant metallicity is Z = 0.015$\pm$0.002 dex. To be able to compare our
metallicity ([Fe/H] = 0.11$\pm$0.03) with that used by other authors we have
derived Z with the approximation given by Eq.~\ref{clusterZ} that corresponds 
to Z = 0.024$\pm$0.002 dex, assuming $Z_{\odot}$ = 0.019 dex. This result is in
agreement with previous metallicity determinations from abundance analysis of
G--type stars by \citet{deokkeun}.

For each star of the sample we have photometrically derived \logl\, and,
adopting the spectroscopic temperatures derived for the abundance analysis,
built the Hertzsprung--Russell (HR) diagram of the cluster around the turn--off
point. We have fitted isochrones by \citet{girardi2002} to our HR diagram to 
derive age and metallicity (\logt\, = 8.77$\pm$0.1 dex; 
Z = 0.030$\pm$0.007 dex). This result is in agreement with that obtained by 
\citet{deokkeun} fitting YREC isochrones to a photometrically--derived main 
sequence of the cluster. We confirm the discrepancy in metallicity found by 
\citet{deokkeun}. Within the errors, our age determination agrees with earlier 
determinations in the literature \citep[\logt\, = 8.85$\pm$0.15,][]{gonzalez}. 

We have plotted the abundances of the elements analysed in most of the stars 
of our sample as a function of \Teff\, and \vsini. We have not found any clear
abundance trend with respect to \Teff\, and \vsini\, for the A-- and F--type
stars. However, we have found several trends between abundances and \vsini\,
for Am stars. Correlations are present for the elements that characterise
the Am peculiarities. Only Sc and Ti do not show this trend. With increasing
\vsini\, the peculiarities tend to weaken, and the abundances approach the 
mean of the other A-- and F--type stars. This is the first real observational 
evidence of a direct connection between abundance anomalies or diffusion 
processes and rotational velocity in Am stars. 

The Praesepe cluster could be peculiar in this sense because it contains a 
high number of Am stars. This fact, combined with the advanced age of the 
cluster, makes it possible to have many Am stars close to the turn--off point. 
Since diffusion is a slow process, in this cluster, more than in others, the 
peculiarities are evident and well--characterised by correlations between 
abundance anomalies and rotational velocity. Similar relationhips may be less 
obvious in younger clusters. It will clearly be of interest to check this 
possibility.

The trends obtained in this work are in disagreement with those predicted 
by \citet{charbonneau}, as according to their models meridional circulation 
does not not affect abundance peculiarities. Recently 
\citet{richer2000} and later \citet{talon} have calculated diffusion models 
with a more accurate treatment of the rotational mixing. Their predicted 
abundances for stars with various rotational velocities are qualitatively in 
agreement with observed properties of our sample of Am stars. Detailed models, 
with a full treatment of diffusion and rotational mixing, calculated
individually for each Am star of our sample, would be necessary for a more 
substantial comparison between models and observed abundances.

We have derived M/M$_\odot$ and fractional age ($\tau$) for each of the stars of
our sample and plotted the abundances as a function of M/M$_\odot$. No
correlation was found for any of the three spectral groups considered here, as
predicted by \citet{richer2000}. 

We have searched for correlations between abundances of the elements analysed
in most of the stars and the Fe abundance. For normal A-- and F--type stars we 
have not found any correlation, while for Am stars strong correlations are 
obtained for all the peculiar elements. The behavior of the abundances of the 
elements which are peculiar in Am stars, as a function of the Fe abundance, is 
the same as that found for \vsini. We conclude that there is a strong 
connection between the correlations among the abundances, with the 
correlations of abundances with \vsini.
\begin{acknowledgements}
We thank Tanya
Ryabchikova for important help and advice given during this investigation.
LF and OK have received support from the Austrian Science Foundation 
(FWF project P17890-N2). JDL and GAW acknowledge support from the Natural 
Science and Engineering Council of Canada (NSERC). GAW acknowledges support 
from the Department of National Defence Academic Research Programme (DND-ARP).
This paper is based on observations obtained using the SOPHIE
spectrograph at the Observatoire de Haute Provence (France). We acknowledge also
the OPTICON program (Ref number: 2007/011) for the financial support given to 
the observing run.
\end{acknowledgements}
\Online
%
%
\end{document}